\documentclass{article}
\usepackage[a4paper, total={6in, 8in}]{geometry}
\usepackage{jheppub}
\usepackage[utf8]{inputenc}
\usepackage{graphicx} 
\usepackage{amssymb}
\usepackage{amsmath}
\usepackage{xcolor}
\usepackage{comment}
\usepackage{ragged2e}
\usepackage{pifont}
\usepackage{multirow} 
\usepackage{booktabs} 
\usepackage{makecell} 
\usepackage{subcaption}

\usepackage[compat=1.1.0]{tikz-feynman}
\DeclareMathOperator{\Tr}{Tr}
\date{December 2024}

\abstract{

   In theories with extended scalar sectors the lightest new scalar degree of freedom might be accessible at colliders. Going beyond simplified models, such a theory can be described in a gauge-invariant and agnostic way via an EFT with a non-linearly realized electroweak symmetry. In this extended HEFT, depending on the $SU(2)$ nature of the new scalar in the UV, operators will be suppressed by different powers of a heavy mass scale. 
We use dimensional analysis to systematically evaluate expected hierarchies between Wilson coefficients, leading to structural relations between potential LHC observables, such as di-boson resonances, tau pair production or the di-photon channel. Once future collider data reveals a hint of a new scalar field, it can be fitted to this extended HEFT and such a structural analysis will help interpret it with respect to possible UV models, circumventing the need to individually test each possible model on the data or to fix the $SU(2)$ representation of the new scalar beforehand. For illustration, the framework is applied to the tentative 95 GeV resonance. In addition to its usefulness for collider physics, the extended HEFT can also be beneficial for low-energy observables, allowing to describe new scalars in an agnostic way.}

\begin{document}

\title{
Characterizing LHC-Resonances in extended HEFT:
information on the nature of extended scalar sectors}

\author[a,e]{Giorgio Arcadi,}
\author[a,b,c]{David Cabo-Almeida,}
\author[d]{Florian Goertz,}
\author[d]{Maya Hager}

\affiliation[a]{Dipartimento di Scienze Matematiche e Informatiche,
Scienze Fisiche e Scienze della Terra, Universita degli Studi di Messina,
Viale Ferdinando Stagno d’Alcontres 31, I-98166 Messina, Italy}
\affiliation[b]{Institut f\"ur Theoretische Physik, Georg-August-Universit\"at G\"ottingen, Friedrich-Hund-Platz 1, 37077 G\"ottingen, Germany}
\affiliation[c]{Departament de Física Quàntica i Astrofísica, Universitat de Barcelona, Martí i Franquès 1, E08028 Barcelona, Spain}
\affiliation[d]{Max-Planck-Institut f\"ur Kernphysik, Saupfercheckweg 1, 69117 Heidelberg, Germany}
\affiliation[e]{INFN Sezione di Catania,
Via Santa Sofia 64, I-95123 Catania, Italy}

\emailAdd{giorgio.arcadi@unime.it}
\emailAdd{david.caboalmeida@uni-goettingen.de}
\emailAdd{florian.goertz@mpi-hd.mpg.de}
\emailAdd{maya.hager@mpi-hd.mpg.de}

\maketitle

\section{Introduction}
With less than 10\% of the designated total LHC data analyzed so far, there remains the realistic possibility that a collider-accessible field beyond the standard model (SM) could be hiding around the corner. In fact, such new TeV-scale states appear naturally in various motivated SM extensions, including models of baryogenesis,  solutions to the hierarchy problem, and in the form of mediators to (thermal-relic) dark matter. 
In this article, reinforced by tentative hints of scalar resonances at the LHC, we present a versatile and theoretically consistent framework to incorporate new spin-0 fields into a effective field theory (EFT) in a generic way, allowing for the possibility that they originate from a non-trivial electroweak representation. We address how, fitting to such an extended Higgs Effective Field Theory (eHEFT), where the particle content of the HEFT \cite{Feruglio:1992wf,Bagger:1993zf,Koulovassilopoulos:1993pw,Grinstein:2007iv,LHCHiggsCrossSectionWorkingGroup:2016ypw,Buchalla:2012qq,Alonso:2012px,Brivio:2013pma,Falkowski:2019tft,Cohen:2020xca} is extended to include an additional (pseudo-)scalar $S$, the structure of excesses can be generically translated to information on its $SU(2)$ nature -- without the need to test different models individually. In this framework, gauge invariance is ensured, making it richer than simplified models \cite{Abdallah:2015ter}, by non-linearly realizing the electroweak symmetry, thus allowing for a systematic and consistent comparison of operators contributing to the collider phenomenology of $S$. This is of particular importance when an excess is observed in more than one channel, as a correlation of signal strengths can give important insights for instance into how the coupling to fermions compares to the coupling to gauge bosons which in turn is intrinsically related to the $SU(2)$ structure of $S$. Employing symmetry and power-counting arguments, the eHEFT analysis extrapolates which properties a UV completion must possess to have caused certain signals, making it a powerful tool in the context of diagnosing anomalies in a generic way.\footnote{See also \cite{Goertz:2017gor} for a related analysis in the context of constraining new-physics masses.} Instead of carrying out a statistical fit for each possible model individually, it is sufficient to match the UV model of interest to the eHEFT Lagrangian and see under which conditions the found relations between coefficients can be reproduced. The framework thus works in two directions; one can start from the coefficients obtained in the best-fit procedure and infer at which canonical dimension the corresponding operators should be generated, using it as a guide for model-building, or one can start with a particular model in mind and see whether the best-fit values for the matched coefficients are in conflict with consistency constraints for the chosen model. Fitting via the eHEFT is economical; several model incarnations can be matched to the same fit, significantly increasing analysis efficiency. In this paper we focus on models containing only one new collider-accessible scalar, but the framework can straightforwardly be extended to allow for more than one in full generality (see also \cite{Arcadi:2024mli}). Moreover, it can also be applied to low-energy observables when seeking to describe new scalars in an agnostic way -- the corresponding analysis will be left for future work.

Analyzing anomalies via the eHEFT will become particularly useful once this or next generation of colliders finds a new signal which then has to be interpreted. While the SM Higgs seems to behave as a $SU(2)$ doublet to good approximation, making Standard Model Effective Field Theory (SMEFT) a good match, for emerging new scalars their electroweak nature is not clear. The eHEFT is therefore well-suited to agnostically describe new physics. Until a significant signal has been found, to illustrate the framework, we apply it to the persisting 95 GeV resonance, which is hinted at by excesses in the di-photon \cite{CMS:2018cyk,CMS:2024yhz,ATLAS:2024bjr} and di-tau channel \cite{CMS:2022goy} from the ATLAS and CMS collaborations, as well as the excess in the di-bottom channel reported by the LEP experiment \cite{LEPWorkingGroupforHiggsbosonsearches:2003ing}. It is one of the most consistent anomalies and has been interpreted in a multitude of papers for a multitude of models \cite{Abdelalim:2020xfk,Cao:2016uwt,Cao:2019ofo,Cline:2019okt,Fox:2017uwr,Liu:2018xsw,Wang:2018vxp,Liu:2018ryo,Choi:2019yrv,Kundu:2019nqo,Bonilla:2023wok,Escribano:2023hxj,Cacciapaglia:2016tlr,Crivellin:2017upt,Haisch:2017gql,Biekotter:2019kde,Biekotter:2020cjs,Heinemeyer:2021msz,Biekotter:2021ovi, Biekotter:2021qbc,Biekotter:2022jyr,Biekotter:2022abc,Li:2022etb,Iguro:2022dok,Iguro:2022fel,Biekotter:2023jld,Banik:2023ecr,Cao:2023gkc,Arcadi:2023smv,Biekotter:2023oen,Azevedo:2023zkg,Belyaev:2023xnv,Aguilar-Saavedra:2023tql, Belyaev:2024lah, Cao:2024axg,Lian:2024smg,Benbrik:2024ptw,Arhrib:2024zsw,Bhatnagar:2025jhh,Ashanujjaman:2023etj,Krishnan:2025haa,Arcadi:2025grl} over the years, highlighting how useful a systematic and unified analysis could be. 

The article is structured as follows. In Sec.\,\ref{sec:Framework}, the eHEFT framework is set up, the power counting explained and a step-by-step guide laid out. Then, in Sec.\,\ref{sec:95GeVres} we apply the method to the 95 GeV resonance as an example. This involves both fitting the EFT parameters in a model-independent way, as well as interpreting the best-fit results and comparing them to predictions of specific classes of models, showcasing how different models can be covered by a single statistical fit. We lay out the properties an ideal model should have to explain the 95 GeV excess in Sec.\,\ref{sec:betterModel}. Ultimately, in Sec.\,\ref{sec:outlook} we give an outlook to the expected behavior of classes of models for a wide mass range of $[100,1000]$ GeV and explain in generality how the ratio between signal strengths can give insights into the $SU(2)$ nature of the scalar resonance. In Sec.\,\ref{sec:conclusion} we conclude and comment on possible extensions.
\section{Framework}
\label{sec:Framework}
Additional scalars furnish an attractive extension of the SM, since they can be relevant for many open questions, such as for explaining the baryon-asymmetry of the universe or for the enigma of dark matter (DM), see \cite{Robens:2025nev} for a recent overview on extended scalar sectors. Even when they are not DM themselves, they can still act as a mediator between a dark sector and the SM, as has been explored in \cite{Alanne:2017oqj,Alanne:2020xcb,Arcadi:2024mli} employing EFT methods. Here, we focus on their interaction with the SM fields and the possibility of detecting them at colliders, while remaining agnostic about other aspects. We consider scalar fields that transform as singlets under $SU(3) \times U(1)_{\rm EM}$ but are allowed to belong to non-trivial representations under $SU(2)$ in the UV. At low energies, when additional modes are integrated out, the scalars can be implemented generically in a gauge invariant way in the phase of the EW symmetry being realized non-linearly, as is commonly done in HEFT for the case of the 125\,GeV Higgs boson, see \cite{LHCHiggsCrossSectionWorkingGroup:2016ypw} for an overview. Here, we extend the HEFT approach by adding an additional (pseudo)scalar to the low energy field content.

\paragraph{Lagrangian} The eHEFT Lagrangian we consider for the Higgs and an additional (pseudo)scalar field $S$ at next to leading order (NLO) reads (for the power counting see below)
\begin{align}
{\cal L} \supset &\frac 1 2 \sum_{\phi=\mathcal{S},h} \partial_{\mu} \phi \partial^{\mu} \phi - \mathcal{F}_{c_s}
+ \frac{v^2}{4} \mathrm{Tr} \left[ \Sigma^{\dagger}
\left( D^{\mu} \Sigma \right) \sigma^{3}\right] \mathrm{Tr} \left[ \Sigma^{\dagger}
\left( D_{\mu} \Sigma \right) \sigma^{3}\right] \mathcal{F}_{c_T}  \nonumber \\
&+ i\frac{v^{2}}{4} \mathrm{Tr} \left[\Sigma^{\dagger}
\left( D^{\mu} \Sigma \right) \sigma^{3} \right] \left( \partial_{\mu} \mathcal{S}\, \mathcal{F}_{S} \right) + \frac{v^{2}}{4} \mathrm{Tr} \left[ \left( D_{\mu} \Sigma \right)^{\dagger}
\left( D^{\mu} \Sigma \right) \right] \mathcal{F}_{\kappa}\nonumber \\
& - \frac{v}{\sqrt{2}} \left( \begin{pmatrix} \overline{u_{i,L}} && \overline{d_{i,L}} \end{pmatrix}
\Sigma \begin{pmatrix} Y_{ij}^{u} u_{j,R} \\ Y_{ij}^{d} d_{j,R} \end{pmatrix} \mathcal{F}_{c_{q}} + \begin{pmatrix} \overline{\nu_{i,L}} && \overline{\ell_{i,L}}  \end{pmatrix} \Sigma \tfrac{1-\sigma_3}{2}Y_{ij}^{\ell} \begin{pmatrix} {\nu_{j,R}} \\ {\ell_{j,R}}  \end{pmatrix} \mathcal{F}_{c_{\ell}} + \mathrm{h.c.} \right) \nonumber \\
&-\frac{1}{16\pi^2}
\left[g^{\prime 2} 
B^{\mu\nu} B_{\mu\nu} \mathcal{F}_{c_{B}}+ g^{2} 
W^{I\mu\nu} W_{\mu\nu}^{I} \mathcal{F}_{c_{W}}+g g^\prime 
B_{\mu \nu} \mathrm{Tr} \left[\Sigma \sigma^3 \Sigma^\dagger W^{\mu\nu}\right] \mathcal{F}_{c_{WB}}
+ g_{s}^{2} 
G^{a\mu\nu} G_{\mu\nu}^{a} \mathcal{F}_{c_G} \right] \nonumber \\
&- \frac{1}{16\pi^2} \left[ g^{\prime 2} 
 B^{\mu\nu} \widetilde{B}_{\mu\nu} \mathcal{F}_{\tilde{c}_{B}} + g^{2} 
 W^{I\mu\nu} \widetilde{W}_{\mu\nu}^{I} \mathcal{F}_{\tilde{c}_{W}}+g g^\prime 
 \widetilde{B}_{\mu \nu} \mathrm{Tr} [\Sigma \sigma^3 \Sigma^\dagger W^{\mu\nu}] \mathcal{F}_{\tilde{c}_{WB}}
+ g_{s}^{2}  G^{a\mu\nu}\widetilde{G}_{\mu\nu}^{a} \mathcal{F}_{\tilde{c}_{G}} \right]\, \nonumber \\
&+ ... \quad ,
\label{Eq:eDMEFT_Lagrangian}
\end{align} 
where the ... are added to account for additional operators which are either not phenomenologically relevant for the here-considered collider signatures \cite{Herrero:2020dtv,Dawson:2023oce} or sub-leading, see below. We refer to \cite{Alonso:2012px} for an extended list for a CP-even scalar, and \cite{Gavela:2014vra} for a CP-odd scalar for the original HEFT case. Above, $f_{L,R}^{(j)}=(1\mp\gamma^{5})f^{(j)}/2$ denotes the chiral SM fermions, $Y^f_{ij}$ are the SM Yukawa matrices, and the $\mathcal{F}_C$ operators collect all scalar combinations of $h$ and $\mathcal{S}$, which are both $SU(2)_L$ singlets in the low energy EFT. 
\begin{align}
\mathcal{F}_C \equiv  \mathcal{F}_C (h, S) \equiv  \sum_{i=0}^2 \sum_{j=0}^{2-i}\, C_{i,j}\, h^i  {\cal{S}}^{j}  + \mathcal{O}((h,S)^3)\, ,
\end{align}	
 generalizing the $(h/v)^n-$polynomials of HEFT \cite{LHCHiggsCrossSectionWorkingGroup:2016ypw}. We explicitly show terms containing up to two scalars, which are relevant for our phenomenology. 
 Gauge invariance is built into the Lagrangian via the Goldstone matrix
\begin{equation}
    \Sigma(x) = e^{i \sigma^j G^j(x)/v}
\end{equation}
which contains the EW Goldstones $G^j$, the Higgs vacuum expectation value (vev) $v$ and the Pauli matrices $\sigma_i$. The Goldstone matrix transforms as
\begin{equation}
    \Sigma(x) \rightarrow e^{i \varphi^j_L \sigma^j/2} \Sigma e^{-i \varphi_Y \sigma^3/2}
\end{equation}
under the SM gauge symmetry. Its covariant derivative is given by
\begin{equation}
    D_\mu \Sigma \equiv \partial_\mu \Sigma - i \frac{g}{2} \sigma^a W_\mu^a \Sigma + i \frac{g^\prime}{2} B_\mu \Sigma \sigma^3.
\end{equation}
 The last two lines of Eq.\,\eqref{Eq:eDMEFT_Lagrangian} are of phenomenological importance for the decays into $\gamma \gamma$, $\gamma Z $ and $gg$ which are otherwise only realized via fermion or EW gauge boson loops, and a loop factor has been added to the operators, since they cannot be tree-generated \cite{Arzt:1994gp}. The operators $ \mathcal{O}_{WB} = g g^\prime  B_{\mu \nu} \Tr[\Sigma \sigma^3 \Sigma^\dagger W^{\mu\nu}] \mathcal{F}_{c_{WB}}$ and $\mathcal{O}_{C_T}=\mathrm{Tr} \left[ \Sigma^{\dagger}
 \left( D^{\mu} \Sigma \right) \sigma^{3}\right] \mathrm{Tr} \left[ \Sigma^{\dagger}
 \left( D_{\mu} \Sigma \right) \sigma^{3}\right] \mathcal{F}_{c_T}$ can contribute considerably to the oblique EW-precision parameters and therefore be tightly constrained, but only if the underlying field has a vev, which might not be the case for the $S$ embedding, while the Higgs sector should be custodially protected. Accordingly, we assume $\mathcal{F}_{c_T}$ and $\mathcal{F}_{c_{WB}}$ to contain no Higgs fields and at least one $S$. Moreover, for simplicity the $\mathcal{F}$ functions of the last two lines should contain at least one physical field in order to keep canonically normalized kinetic terms. Lifting this assumption would only have a subleading impact on the analysis. If $S$ is CP-odd (and CP is conserved), $c^S_{B,W,WB,G} = 0$ and vice versa. In this case, the derivative operator $i\frac{v^{2}}{4} \mathrm{Tr} \left[\Sigma^{\dagger}
\left( D^{\mu} \Sigma \right) \sigma^{3} \right] \partial_{\mu}\mathcal{S} \,\mathcal{F}^{(\geqslant1h)}_{S}$ becomes relevant, as it generates the $Z h S$ interaction characteristic of a CP-odd scalar \cite{Bauer:2016zfj}, an interaction that is not allowed for a CP-even scalar. At least one Higgs has to be contained in $\mathcal{F}_S$, otherwise the operator becomes redundant.\footnote{Notice that, employing integration by parts and equations of motion, any terms not mixing different scalars and  following the same structure can be rewritten in terms of fermionic currents. All possible non-redundant derivative couplings involving one derivative are instead accounted for in $i\frac{v^{2}}{4} \mathrm{Tr} \left[\Sigma^{\dagger}
\left( D^{\mu} \Sigma \right) \sigma^{3} \right] \partial_{\mu}\mathcal{S} \,\mathcal{F}_{S}$.}

\paragraph{Power-Counting}
SMEFT \cite{Buchmuller:1985jz,Grzadkowski:2010es}, a linear EFT, employs canonical mass dimension for its power-counting. In contrast, HEFT uses power-counting by chiral dimension $D_c(F) \equiv [F]$, where (see, e.g. \cite{Alonso:2012px,Buchalla:2013eza,LHCHiggsCrossSectionWorkingGroup:2016ypw,Brivio:2013pma,Brivio:2025yrr})
\begin{equation}
    [X_\mu, \phi] = 0, \quad [\partial,y,g] = 1, \quad [\psi] = 1/2,
\end{equation}
so gauge fields and scalars have chiral dimension 0, derivatives, gauge and Yukawa couplings have chiral dimension 1, while fermions have chiral dimension 1/2. Operators with more scalars are not generically suppressed in the HEFT. The chiral dimension of an operator is related to its loop order $L$ by $D_c = 2 L + 2$. In the eHEFT of Eq.\,\eqref{Eq:eDMEFT_Lagrangian} the first three lines contain operators of chiral dimension two, corresponding to the leading order terms, while the last lines contain operators of chiral dimension four, they are loop-suppressed. NLO is then described by taking into account the $D_c=4$ terms as well as loops arising from the $D_c=2$ terms. Expanding in chiral dimension is particularly useful in non-linear EFTs when the deviation of couplings cannot be constrained to be $\ll 1$, as for instance in strongly coupled theories, such as composite Higgs frameworks at small decay constant $f\sim v$, see \cite{Contino:2010rs,Bellazzini:2014yua,Panico:2015jxa,Goertz:2018dyw} for reviews. Accordingly, the Higgs boson does not combine with the EW Goldstones to a proper $SU(2)_L$ doublet and the cutoff of the pure HEFT resides at $\Lambda \sim 4\pi v$.

Here, we consider the scalar $S$ to reside at most at the $\sim 1$\,TeV scale and thereby being LHC accessible -- with additional components, should it originate from a non-trivial $SU(2)$ multiplet, being heavier by $\delta m \gtrsim {\rm(a\ few)} \times v$ and integrated out, leading to the EFT of \eqref{Eq:eDMEFT_Lagrangian}. We thus assume at least a moderate hierarchy between the mass $m_S$ and the EFT suppression scale $\Lambda$, which for the different operators could be set by a $SU(2)$ symmetric mass of states beyond the EFT or by $\Lambda \sim 4 \pi v$ in the non-linear HEFT limit, $m_S \lesssim {\rm TeV}  \ll \Lambda$. Both the HEFT limit and the (approximate) linear limit can be captured by the eHEFT, with the full linear limit corresponding to $S$ being a full EW singlet in the UV while the Higgs belongs to a doublet. In this case, full decoupling $\Lambda\to \infty$ can be achieved.
In the following we consider a hierarchical limit such that the sub-set of the chiral--dimension $D_c=4$ terms not explicitly included in \eqref{Eq:eDMEFT_Lagrangian} features indeed an additional suppression by a heavy mass (or the operators do not give a leading contribution to scalar resonance observables at the LHC).
We note that this limit would no longer hold if the new field were to obtain a large vev $v_S$ and therefore a hierarchy $v_S/\Lambda \gtrsim 1$.

\paragraph{Procedure} 
\begin{figure}[t!]
    \centering
    \includegraphics[width=0.7\linewidth]{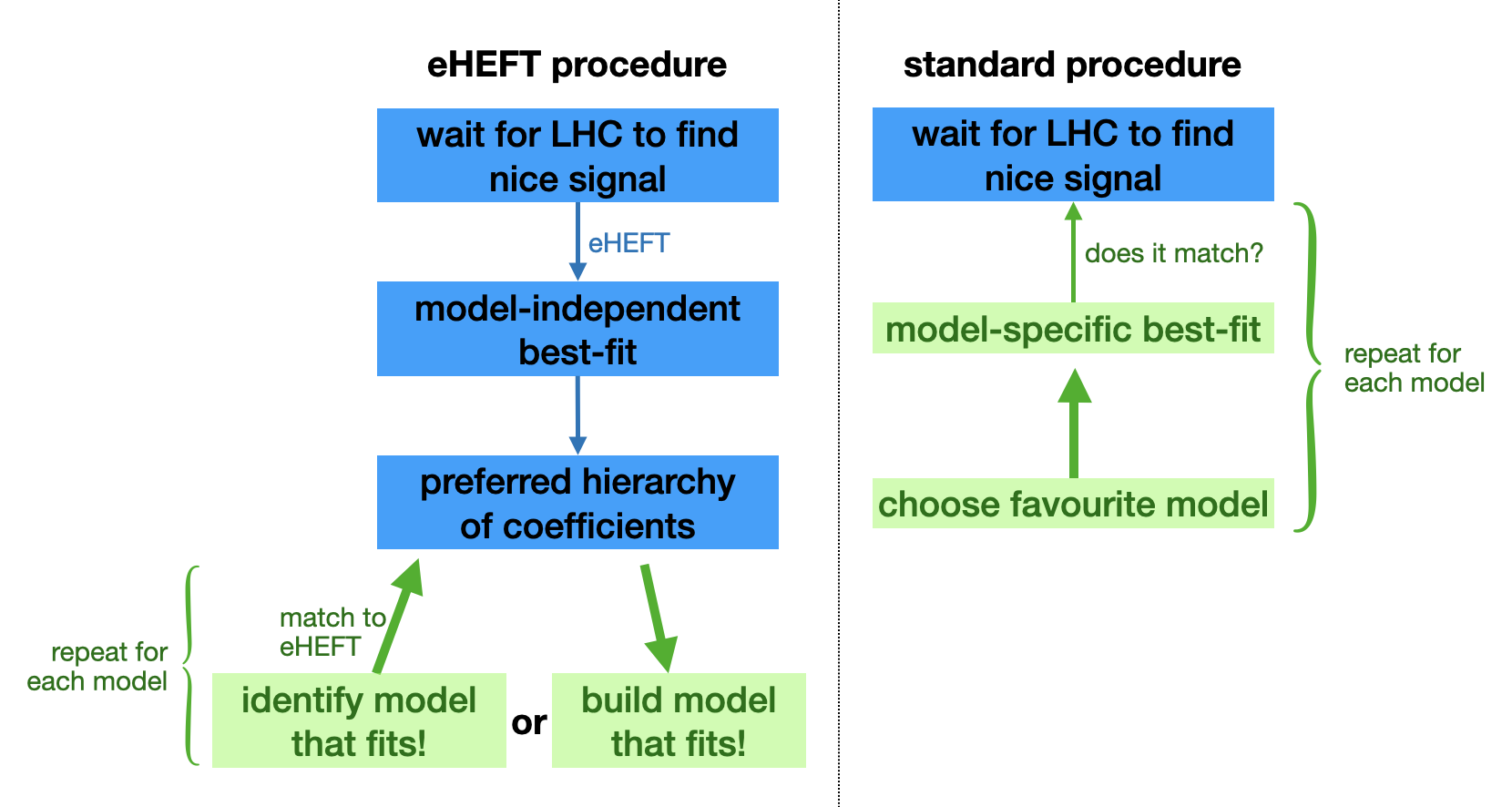}
    \caption{Schematic Illustration of the difference between the eHEFT approach and the standard way of analyzing models of scalar extensions. Matching to the eHEFT is more economical than redoing the statistical analysis for each model, and once the hierarchy of coefficients is determined it also acts as a guiding principle that can be used to build a fitting model, therefore making it richer than the standard procedure.}
    \label{fig:schematicEHEFT}
\end{figure}
In the following, we call
\begin{equation}
    c_W \equiv (c_W)_{0,1}, \quad c_B \equiv (c_B)_{0,1}, \quad c_{WB} \equiv (c_{WB})_{0,1}, \quad c_G \equiv (c_G)_{0,1}, 
\end{equation}
collectively $c_X$, and, for the operators that include factors of $v$,
we define 
\begin{equation}
    \kappa^\prime \equiv v \, \kappa_{0,1}, \quad c_q^\prime \equiv v \, (c_q)_{0,1} , \quad c^\prime_\ell \equiv v \, (c_\ell)_{0,1} , \quad c_T^\prime \equiv v \, (c_T)_{0,1}  \, ,
\end{equation} 
in order to preferentially work with dimensionless coupling parameters. By absorbing the scale dependence into the coefficients we remain agnostic about the origin of the corresponding operators during the analysis.\footnote{Note that in case an operator is induced by a new vev $v_s$, the matched coefficient will contain a ratio of vevs.} A preference for a low value of a specific coefficient will then hint at a larger canonical operator dimension it corresponds to (in the limit of sufficient scale separation)  and vice versa. Since we are mainly concerned about trilinear interactions containing one $S$ and two SM fields, we have chosen to normalize the couplings by only one power of the vev. If the process of interest were to involve two insertions of $S$, one could for instance redefine $\kappa^{\prime\prime} \equiv v^2 \kappa_{0,2}$ to obtain a dimensionless coupling parameter. By keeping the SM Yukawa matrices in the third line of Eq.\,\eqref{Eq:eDMEFT_Lagrangian}, flavor alignment is assumed. It is however straight-forward to modify this assumption; one would simply need to normalize the $c_q, c_\ell$ couplings by the Yukawas. To analyze the production cross section and decay widths, we consider the tree-level contributions of the EFT Lagrangian to the decays of $S$ into SM fields, as well as the loop-mediated decays into $WW, \, ZZ, \, Z \gamma, \, \gamma \gamma, \, gg$ and the off-shell decays into $W W^\ast,\, Z Z^\ast$. The analytical expressions for the decay widths are given in Appendix \ref{app:decayWidths}. Then,
\begin{itemize}
    \item[1.   ] choose LHC signals to fit,
    \item[2.   ] scan over the coefficients in suitable range,
    \item[3.   ] calculate the $\chi^2$ value,
    \begin{equation}
        \chi^2 = \sum_i\frac{\left( \mu_i - \mu_i^\text{exp}\right)^2}{\left(\Delta \mu_i^\text{exp} \right)^2},
    \end{equation}
    where $i$ sums over all relevant channels,
    $\mu$ is the theoretical signal strength, $\mu_i^\text{exp}$ the central value of the measurement and $\Delta \mu_i^\text{exp}$ its 1$\sigma$ uncertainty,
    \item[4.   ] find parameter space for the coefficients that matches the observed excess,
    \item[5a.   ] infer what this means for constructing a UV-model, 
    \item[5b.   ]or choose a specific UV model and see if it can reproduce the hierarchies of the coefficients.
\end{itemize}
Before the last step, there are no assumption about the model, as long as it is well described by the eHEFT Lagrangian 
\eqref{Eq:eDMEFT_Lagrangian}. Fig.\,\ref{fig:schematicEHEFT} schematically shows the difference between this approach and the standard approach where the fitting has to be carried out for each model individually. Thus, using the eHEFT analysis is more efficient and can additionally give insights into what an ideal model should feature as hierarchies between the coefficients, working in two ways. In the next section, we will illustrate the framework by applying it to the tentative 95 GeV resonance, showing in detail how the statistical analysis can be carried out.

\section{95 GeV Resonance}
\label{sec:95GeVres}
As a benchmark example, we test our EFT framework against current collider anomalies, focusing on the persistent hint of a resonance around $95$ GeV. The CMS Collaboration first reported an excess in the di-photon channel with a local significance of $2.9\sigma$ at $m_{\gamma\gamma}\approx95\,\mathrm{GeV}$~\cite{CMS:2018cyk,CMS:2024yhz}. An independent ATLAS search in the same final state observed a local significance of $1.7\sigma$~\cite{ATLAS:2024bjr}. Combining both results yields a signal strength $\mu_{\gamma\gamma}^{\mathrm{CMS}+\mathrm{ATLAS}} = 0.24_{-0.08}^{+0.09}$~\cite{Biekotter:2023oen}. This hint is further supported by a CMS search in the di-tau channel, reporting a signal strength $\mu_{\tau \tau} = 1.2\pm0.5$~\cite{CMS:2022goy}, consistent with the earlier observation of an excess in the di-bottom final state by the LEP experiments in the process $e^+e^-\to Z\mathcal{S}\to b\bar{b}$ at $m_{bb}\approx98\,\mathrm{GeV}$, corresponding to a local significance of $2.3\sigma$ and a signal strength $\mu_{bb}=0.117\pm0.06$~\cite{LEPWorkingGroupforHiggsbosonsearches:2003ing,Cao:2016uwt}.

These correlated excesses across different final states motivate the interpretation of the signal as a new scalar resonance. Within our eHEFT framework, such a state can be parametrized by a scalar field $\mathcal{S}$ whose interactions with SM fields are described by the effective Lagrangian in Eq.(\ref{Eq:eDMEFT_Lagrangian}). 

In the following, we focus on a scenario in which the new scalar is CP-even. Moreover, the production of a 95 GeV SM-like scalar is dominated by gluon fusion and we adopt the same hypothesis in our study of the new scalar $S$. Although this mechanism is the primary production mode for spin-0 states in the SM, extended Higgs sectors can feature enhanced couplings to bottom quarks, potentially increasing the contribution from $b\bar{b}$ fusion. Nevertheless, following the experimental analyses, we focus on gluon fusion. We note that dijet final states have also been explored by CMS with masses as low as 50\,GeV, though current limits near 95\,GeV remain weak due to large QCD backgrounds \cite{CMS:2018pwl}. Under this assumption, the gluon-fusion production cross section can be written as \cite{Spira:1995rr} 
\begin{equation}
\sigma_{gg}=\frac{8\,\pi^2}{M_S\,s}\Gamma(S\to gg)\frac{d\mathcal{L}}{d\tau_S},
\end{equation}
where the Drell–Yan variable is defined as $\tau_S=M_S^2/s$, with $s=(13\,\mathrm{TeV})^2$, and $\frac{d\mathcal{L}}{d\tau_S}$ denotes the parton luminosity factor accounting for the proton PDFs,
\begin{align}
\frac{d\mathcal{L}}{d\tau} & =\int_{\tau}^1 \frac{d x}{x} g(x) g\left(\frac{\tau}{x}\right)\,.
\end{align}
Due to the assumption of gluon-fusion dominance, the signal strengths for the $\tau\tau$ and $b\bar{b}$ final states simplify to
\begin{equation}
\mu_{\tau \tau}=\frac{\sigma_{gg}\,BR(S\to \tau\tau)}{\sigma_{gg}^\mathrm{SM-like}\,BR(S\to \tau\tau)_\mathrm{SM-like}}=\frac{\Gamma(S\to gg)\,BR(S\to \tau\tau)}{\Gamma(S\to gg)_\mathrm{SM-like}\,BR(S\to \tau\tau)_\mathrm{SM-like}},
\end{equation}
\begin{equation}
\mu_{b b}=\frac{\sigma_{Z S} \, B R(S \rightarrow b b)}{\sigma_{Z S}^\mathrm{SM-like} \, B R(S \rightarrow b b)_\mathrm{SM-like}}=\frac{{\kappa^\prime}^2 B R(S \rightarrow b b)}{4\, B R(S \rightarrow b b)_\mathrm{SM-like}}
\end{equation}
where ``SM--like'' stands for the value corresponding to a $95\,\mathrm{GeV}$ scalar with SM--like couplings.

For the diphoton channel, we additionally account for vector-boson–fusion (VBF), $V$--associated production (VS), and $t\bar t$ associated production ($ttS$), and define
\begin{equation}
\mu_{\gamma \gamma}=\frac{R_{g g} \sigma_{g g}^\mathrm{SM-like}+R_V \sigma_{V B F+V S}^\mathrm{SM-like}+R_{t t} \sigma_{t t S}^\mathrm{SM-like}}{\sigma_{S}^\mathrm{SM-like}}\,\frac{B R(S \rightarrow \gamma \gamma)}{B R(S\rightarrow \gamma \gamma)_\mathrm{SM-like}},
\end{equation}
where $R_{gg}$, $R_V$, and $R_{tt}$ encode the relative change in gluon-fusion production, VBF and VH production, and $t \bar t$ associated production, respectively,
\begin{align} 
R_{g g}&=\frac{\Gamma(S \rightarrow g g)}{\Gamma(S\rightarrow g g)_\mathrm{SM-like}}, \\ 
R_V & =\left(\frac{\kappa^\prime}{2}\right)^2,\\
R_{tt} & ={c_t^\prime}^2. 
\end{align}
We obtain the gluon-fusion cross section using \texttt{SusHi v1.7.0} at NNLO accuracy~\cite{Harlander:2012pb,Harlander:2016hcx}, and use SM-like reference values for the different production channels of a 95\,GeV scalar resonance, $\sigma_{V B F+V H}^\mathrm{SM-like}(13\,\mathrm{TeV})=10.4\,\mathrm{pb}$, $\sigma_{ttS}^\mathrm{SM-like}(13\,\mathrm{TeV})=1\,\mathrm{pb}$ and $\sigma_{S}^\mathrm{SM-like}=\sigma_{V B F+V H}^\mathrm{SM-like}+\sigma_{ttS}^\mathrm{SM-like}+\sigma_{gg}^\mathrm{SM-like}\simeq 87.7\,\mathrm{pb}$~\cite{LHCHiggsCrossSectionWorkingGroup:2016ypw}, while the decay widths can be obtained from Appendix \ref{app:decayWidths}.
\begin{figure}[b!]
    \begin{tikzpicture}[baseline=10mm]
        \begin{feynman}
            \vertex (j1) {\(g\)};
            \vertex[above=1.5cm of j1] (j2) {\(g\)};

            \vertex at ($(j1)!0.5!(j2) + (1cm,0)$) (k);
            \vertex[right=1cm of k] (l);
            \node[dot] at (k){};
            
            \vertex[left=0.2cm of k] (z) {\(c^\prime_q{,}\,c_G\)};
            
            \vertex[right=1.2cm of k] (z) {\(c_f^\prime\)};
            \vertex at ($(j1) + (3cm,0)$) (m1) {\(f\)};
            \vertex at ($(j2) + (3cm,0)$) (m2) {\(\bar{f}\)};
        
            \diagram* {
            (j1) -- [gluon] (k),
            (j2) -- [gluon] (k),
            (k) -- [scalar,edge label=\(S\)] (l),
            (l) -- [fermion] (m1),
            (m2) -- [fermion] (l),
            };
        \end{feynman}
    \end{tikzpicture}
    \hspace{2mm}
    \centering
    \begin{tikzpicture}[baseline=10mm]
        \begin{feynman}
            \vertex (j1) {\(g\)};
            \vertex[above=1.5cm of j1] (j2) {\(g\)};
            
            \vertex at ($(j1)!0.5!(j2) + (1cm,0)$) (k);
            \node[dot] at (k){};
            \node[right=1cm of k,dot] (l);
            \vertex[left=0.2cm of k] (z) {\(c^\prime_q{,}\,c_G\)};
            
            \vertex[right=1.2cm of k] (z) {\(\kappa^\prime{,}\,c^\prime_f{,}\,c_B{,}\,c_W{,}\,c_{WB}\)};
            \vertex at ($(j1) + (3cm,0)$) (m1) {\(\gamma\)};
            \vertex at ($(j2) + (3cm,0)$) (m2) {\(\gamma\)};
        
            \diagram* {
            (j1) -- [gluon] (k),
            (j2) -- [gluon] (k),
            (k) -- [scalar,edge label=\(S\)] (l),
            (l) -- [boson] (m1),
            (m2) -- [boson] (l),
            };
        \end{feynman}
    \end{tikzpicture}
    \hspace{2mm}
    \centering
    \begin{tikzpicture}[baseline=10mm]
        \begin{feynman}
            \vertex (j1) {\(e^-, q\)};
            \vertex[above=1.5cm of j1] (j2) {\(e^+, q\)};

            \vertex at ($(j1)!0.5!(j2) + (1cm,0)$) (k);
            \vertex[right=1cm of k] (l);
            
            \vertex[left=0.2cm of k] (z) {};
            \vertex[right=1.2cm of k] (z) {\(\kappa^\prime{,}\,c_T^\prime{,}\,c_B{,}\,c_W{,}\,c_{WB}\)};
            \vertex at ($(j1) + (3cm,0)$) (m1) {\(Z\)};
            \vertex at ($(j2) + (3cm,0)$) (m2) {\(S\)};
        
            \diagram* {
            (j1) -- [fermion] (k),
            (k) -- [fermion] (j2),
            (k) -- [boson,edge label=\(Z\)] (l),
            (l) -- [boson] (m1),
            (m2) -- [scalar] (l),
            };
        \end{feynman}
    \end{tikzpicture}
    \centering
    \begin{tikzpicture}[baseline=10mm]
        \begin{feynman}
            \vertex (j1) {\(q\)};
            \vertex[above=2cm of j1] (j2) {\(q\)};

            \vertex at ($(j1)!0.5!(j2) + (1.5cm,0)$) (k);
            
            \vertex[left=0.05cm of k] (a) {\(\kappa^\prime{,}\,c_T^\prime{,}\,c_B{,}\,c_W{,}\,c_{WB}\)};
            \vertex[above=0.5cm of k] (l1);
            \vertex[below=0.5cm of k] (l2);
            \vertex[right=1.2cm of k] (z) {\(S\)} ;
            \vertex at ($(j1) + (3cm,0)$) (m1) {\(q\)};
            \vertex at ($(j2) + (3cm,0)$) (m2) {\(q\)};
        
            \diagram* {
            (j1) -- [fermion] (l2),
            (j2) -- [fermion] (l1),
            (l1) -- [fermion] (m2),
            (l2) -- [fermion] (m1),
            (l1) -- [boson] (l2),
            (k) -- [scalar] (z),
            };
        \end{feynman}
    \end{tikzpicture}
    \centering
    \begin{tikzpicture}[baseline=10mm]
        \begin{feynman}
            \vertex (j1) {\(g\)};
            \vertex[above=2cm of j1] (j2) {\(g\)};

            \vertex at ($(j1)!0.5!(j2) + (1.5cm,0)$) (k);
            
            \vertex[left=0.05cm of k] (a) {\(c^\prime_q\)};
            \vertex[above=0.5cm of k] (l1);
            \vertex[below=0.5cm of k] (l2);
            \vertex[right=1.2cm of k] (z) {\(S\)} ;
            \vertex at ($(j1) + (3cm,0)$) (m1) {\(t\)};
            \vertex at ($(j2) + (3cm,0)$) (m2) {\(t\)};
        
            \diagram* {
            (j1) -- [gluon] (l2),
            (j2) -- [gluon] (l1),
            (m2) -- [fermion] (l1),
            (l2) -- [fermion] (m1),
            (l1) -- [fermion] (l2),
            (k) -- [scalar] (z),
            };
        \end{feynman}
    \end{tikzpicture}
    \caption{Feynman diagrams of the processes relevant to the $95$ GeV scalar resonance, illustrating the relevant Wilson coefficients. The fermionic final states correspond to $f\bar{f} = b\bar{b},\tau\tau$. A dot stands for a vertex, in which loop effects can contribute. For the additional production channels in the last line we do not explicitly show the subsequent $S$ decay.
    } 
    \label{Fig:95GeVDiagrams}
\end{figure}
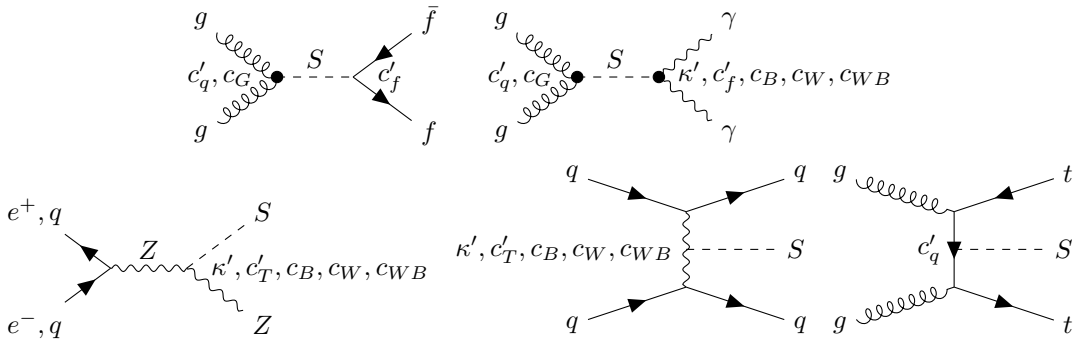
\subsection{Model-Independent Analysis
\label{sec:mod-indAnalysis}}
\begin{figure}[t!]
    \centering
    \begin{subfigure}{0.49\textwidth}
    \includegraphics[width=\linewidth]{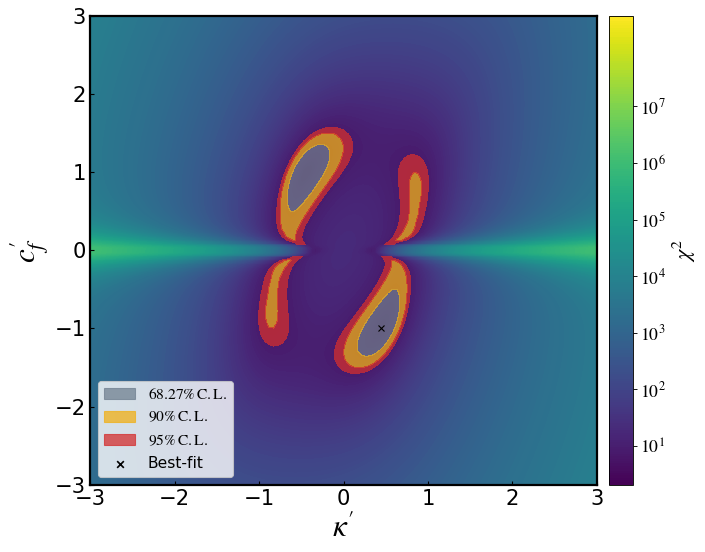}
    \caption{$c^\prime_f,\,\kappa^\prime$ and $\chi^2_{\text{bf}} = 2.1$}
    \label{fig:chisquareUniversal}
    \end{subfigure}
    \hfill
    \begin{subfigure}{0.49\textwidth}
    \includegraphics[width=\linewidth]{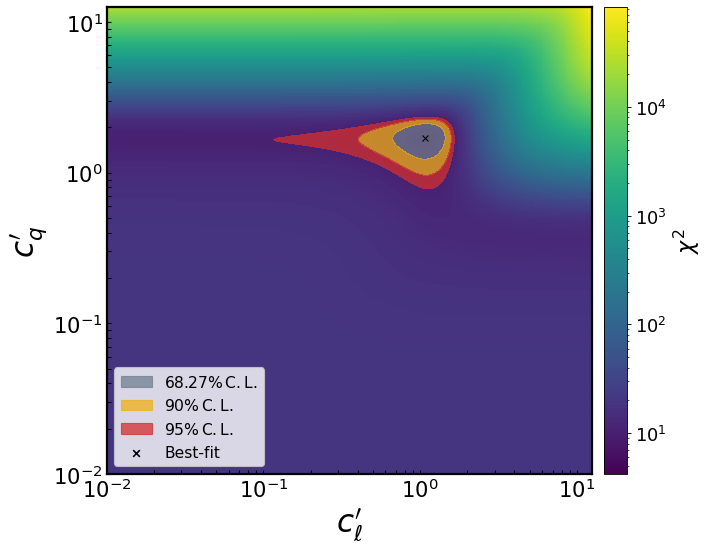}
    \caption{$\kappa^\prime=0,\,c^\prime_q,\,c^\prime_\ell$ and $\chi^2_{\text{bf}} = 4.2$}
    \label{fig:chisquareKappazero}
    \end{subfigure}
    \medskip
    \begin{subfigure}{0.49\textwidth}

    \vspace{4mm}
    \includegraphics[width=\linewidth]{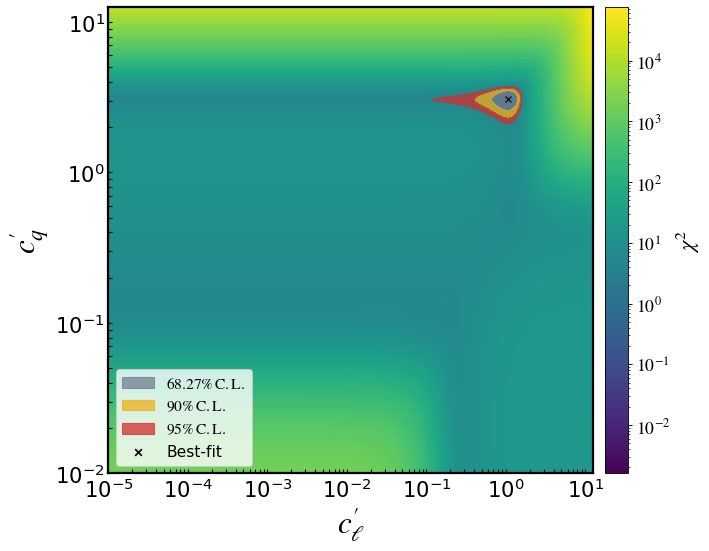}
    \caption{$\kappa^\prime=0.65,\,c^\prime_q,\,c^\prime_\ell$ and $\chi^2_{\text{bf}} = 6.4\cdot10^{-4}$}
    \label{fig:chisquareKappa6}
    \end{subfigure}
    \hfill
    \begin{subfigure}{0.49\textwidth}
    \includegraphics[width=\linewidth]{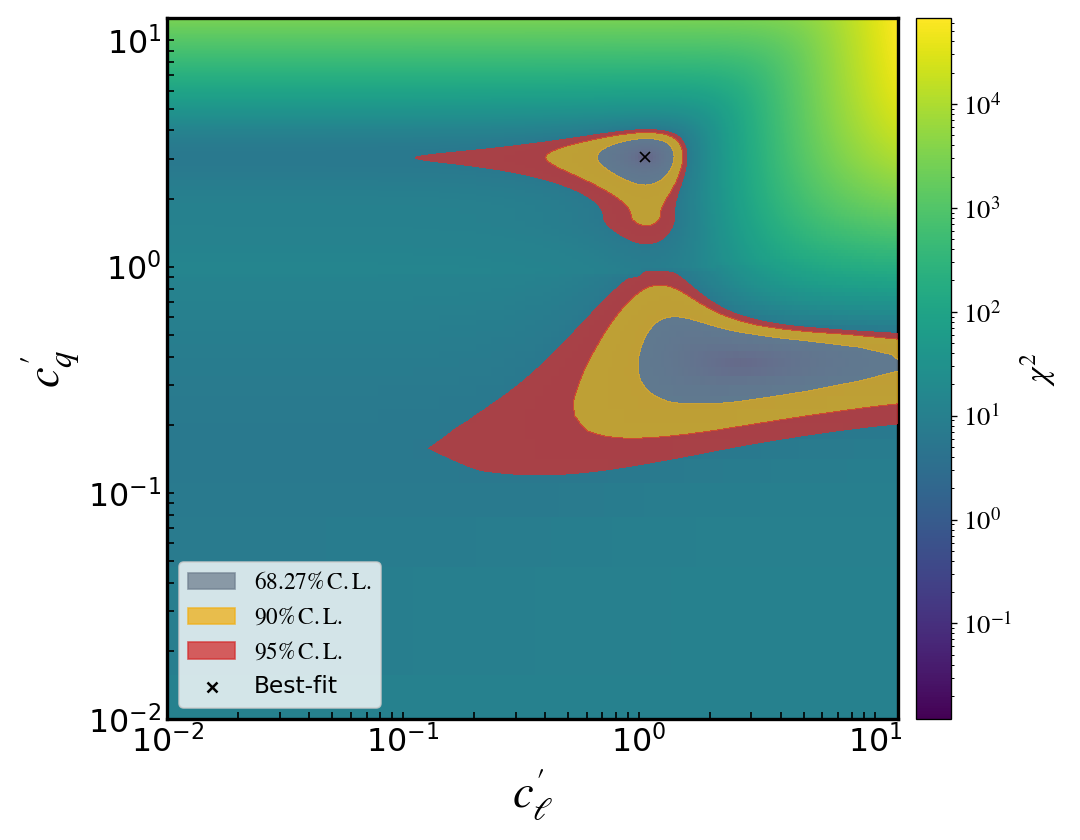}
    \caption{$\kappa^\prime,\,c^\prime_q,\,c^\prime_\ell$ and $\chi^2_{\text{bf}} = 4.4\cdot10^{-6}$}
    \label{fig:chisquareMarginalize}
    \end{subfigure}
    \caption{Results of the $\chi^2$ analysis matching the EFT to the observed excesses at $95$\,GeV. Panel (a) shows the case of a universal fermion coupling $(c^\prime_f,\kappa^\prime)$, while (b) and (c) correspond to independent quark and lepton couplings $(c^\prime_q,c^\prime_\ell)$ with $\kappa^\prime=0$ and $\kappa^\prime=0.65$, respectively. Panel (d) displays the case where $\kappa^\prime$ is marginalized over. The color maps indicate the $\chi^2$ values, with the shaded contours representing the $68.27$\%, $90$\% and $95$\% confidence regions, and the cross marking the best-fit point.}
    \label{fig:chisquare}
\end{figure}
The EFT framework can generate the processes shown in Fig.\,\ref{Fig:95GeVDiagrams}, which are described by the couplings $\kappa^\prime$, $c_T^\prime$, $c^\prime_f$, $c_G$, $c_B$, $c_W$ and $c_{WB}$. In order to reproduce the observed excesses across the three channels, we focus on the scenario where the gluon-fusion production is dominated by SM quarks, and is therefore mainly controlled by the coupling $c^\prime_q$. The coefficient $c_G$ encodes potential loop-induced contributions from heavy new fields to the gluon interactions, while $c_B$, $c_W$ and $c_{WB}$ parametrize analogous effects for the electroweak gauge bosons. For simplicity, at first, we set these loop-induced coefficients, as well as $c_T^\prime$, to zero for the 95 GeV analysis, see Appendix \ref{app:operatorsChDim4} and Appendix \ref{app:decayZZ} for a discussion on the magnitude of their effect over a wide mass range. Consequently, we analyze the case in which $\kappa^\prime$ and $c^\prime_f$ are treated as free parameters. 
The relative sign between $\kappa^\prime$ and $c^\prime_f$ is relevant; therefore, in the initial analysis we scan the dimensionless coefficients in the range $[-4\pi,4\pi]$. For the other parameter combinations considered in the analysis, the results are symmetric under sign changes (see Appendix~\ref{app:couplingRange}), and we therefore restrict the scan to the interval $[0,4\pi]$. In the logarithmic scan, the minimum value is set to $10^{-5}$. If the corresponding operator is generated at canonical mass dimension five ($c_i^\prime \sim v/\Lambda$), see Sec.\,\ref{sec:specificModels} for specific models, this corresponds to $\Lambda \sim 10^4$ TeV, or for a canonical dimension six operator ($c_i^\prime \sim v^2/\Lambda^2$), to $\Lambda \sim 10^2$ TeV. At both scales we would expect the eHEFT description to have broken down if $S$ originates from a non-trivial multiplet, due to the large required separation between the additional components and $S$, and furthermore, it would no longer be possible to generate observable collider excesses when the operators experience such a large suppression, making $c_i \sim 10^{-5}$ equivalent to $c_i \sim 0$ up to negligible effects. 

To quantify the compatibility between the EFT predictions and the observed excesses,\footnote{In this proof of concept we focus on fitting the channels in which the excess was observed, rather than fitting all measured channels.} we perform a $\chi^2$ analysis, evaluating the fit results at the $68.27$\%, $90$\%, and $95$\% confidence levels. We begin by considering the case of a universal fermionic Yukawa coupling, $c^\prime_f$, and determine the corresponding best-fit regions, shown in Fig.~\ref{fig:chisquareUniversal}. We observe that at the $68.27$\% confidences level, solutions with $\kappa^\prime = 0$ remain allowed. Even though the diphoton and dibottom channels are not reproduced, the fit cannot exclude this scenario with the present experimental statistics. Nevertheless, to relax the assumption of a universal fermionic coupling, we fix the parameter $\kappa^\prime$ to specific values and distinguish between the quark and lepton couplings, $c^\prime_q$ and $c^\prime_\ell$. The choice of $\kappa^\prime$ is nontrivial, as it directly affects the diphoton channel. 

We first fix $\kappa^\prime = 0$, obtaining the allowed coefficient regions shown in Fig.\,\ref{fig:chisquareKappazero}. Since the diphoton channel cannot be reproduced in this case, the fit to the other two channels becomes more constrained, resulting in a reduced viable parameter space.

\begin{figure}
    \centering
    \includegraphics[width=0.49\linewidth]{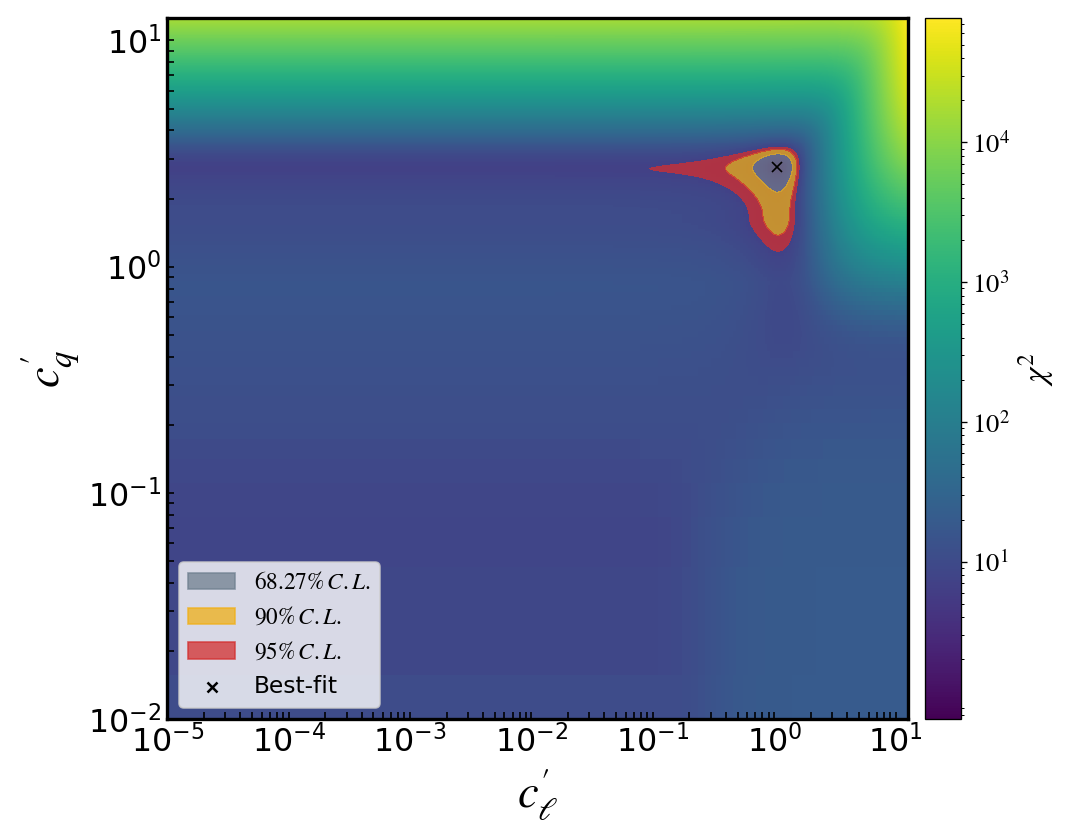}
    \hfill
    \caption{Results of the $\chi^2$ analysis matching the EFT to the observed excesses at $95$\,GeV for $\kappa^\prime\lesssim0.5,\,c^\prime_q,\,c^\prime_\ell,\,c_X = 0.24$ TeV$^{-1}$ (motivated by model (I) in Sec.\,\ref{sec:specificModels}), where $\chi^2_{\text{bf}} = 0.7$.}
    \label{fig:marginalization_small}
\end{figure}

For the case of nonzero $\kappa^\prime$, we treat $\kappa^\prime$ as a nuisance parameter and marginalize over it (as well as zoom in on a preferred value). As noted before, the relative sign of $\kappa^\prime$ is relevant. Nevertheless, as discussed in Appendix~\ref{app:couplingRange}, the result of the marginalization is symmetric under a sign change of the best-fit value of $\kappa^\prime$. We consider two scenarios. First, in Fig.\,\ref{fig:chisquareMarginalize}, $\kappa^\prime$ is an unconstrained free parameter. The resulting best-fit exhibits two disconnected regions in the $(c_q^\prime,c_\ell^\prime)$ plane. The first, corresponding to higher values of $c_q^\prime$, prefers $\kappa^\prime \simeq [0.0012,0.86]$ at the $90$\% level with best-fit values $\chi^2_\mathrm{bf}=4.4 \cdot 10^{-6}$ for $\kappa^\prime \sim0.65$, $c_q^\prime \sim 3.1$, $c_\ell^\prime \sim 1.1$. The second region, associated with lower values of $c_q^\prime$, yields $\kappa^\prime \simeq [0.87,4.07]$ and is slightly less favored ($\chi_\mathrm{bf}=0.03$), although the difference is not statistically significant. These two solutions therefore reflect distinct scenarios, as illustrated by Fig.\,\ref{fig:chisquareKappa6}, where we fix $\kappa^\prime = 0.65$ and the second region visible in Fig.\,\ref{fig:chisquareMarginalize} disappears. Second, in Fig.\,\ref{fig:marginalization_small} we include the effect of the chiral--dimension four operators. In order not to overshadow its effect, we consider the case in which $\mathcal{O}_\kappa$ does not dominate over $\mathcal{O}_X$, i.e. when all operators are generated at mass dimension five. Therefore, we have chosen to set $c_X = 0.24 \, \mathrm{TeV}^{-1}$ and limit $\kappa^\prime \lesssim 0.5$, both choices consistent with the operators arising at mass dimension five, see Sec.\,\ref{sec:specificModels} for details, where we consider specific $SU(2)$ models. The best-fit region maximizes $\kappa^\prime \rightarrow 0.5$ and yields the best-fit $\chi^2_\mathrm{bf}=0.7$ with $c_q^\prime = 2.75$ and $c_\ell^\prime = 1.05$. Although an additional parameter has been added, the fit gets worse due to the limit on $\kappa^\prime$, since this coupling directly affects two of the three channels included in the fit, see Sec.\,\ref{sec:betterModel} for a discussion on how to allow large $\kappa^\prime$ consistent with the best-fit.

\subsection{Constraining Specific Models}
\label{sec:specificModels}
In this section, we illustrate how in this framework the best-fit can be used to constrain specific models. Instead of redoing the fit for each model incarnation, the UV models can be matched to the eHEFT, which will give predictions for the coefficients fitted in the previous section. In this outlook we focus on comparing simple scalar toy models with vanishing expectation value and to limit the number of parameters allow no terms where $\sim H S$ forms a singlet, which would e.g. lead to mixing with the Higgs. Of course, the analysis can be extended to models where the new scalar obtains a vev or mixes with the Higgs, although one has to track relations between the operator coefficients that might arise from necessary field redefinitions and take into account additional parameters. For instance, in a 2-Higgs-Doublet-Model (2HDM) the number of structures that have to be fitted is increased to include mixing angles and a second vev. There is a rich literature on interpreting the 95 GeV excess in the context of 2HDMs \cite{Cacciapaglia:2016tlr,Crivellin:2017upt,Haisch:2017gql,Biekotter:2019kde,Biekotter:2020cjs,Heinemeyer:2021msz,Biekotter:2021ovi, Biekotter:2021qbc,Biekotter:2022jyr,Biekotter:2022abc,Li:2022etb,Iguro:2022fel,Biekotter:2023jld,Banik:2023ecr,Biekotter:2023oen,Azevedo:2023zkg,Belyaev:2023xnv,Aguilar-Saavedra:2023tql, Belyaev:2024lah, Benbrik:2024ptw,Arhrib:2024zsw,Bhatnagar:2025jhh}, showcasing how beneficial the here-proposed general analysis could be: the best fit is done for classes of models and to compare it to a UV completion it is sufficient to determine the behavior of the coefficients in the 2HDM variant, i.e. determine their scaling and restrict their range to be consistent with UV-model-specific constraints -- making it much more economical than redoing the analysis for each variation and, more importantly, enabling a unified and model-independent comparison of different $SU(2)$ representations for the new scalar, rather than restricting the analysis to a specific doublet realization.

Starting with a $SU(2)$ singlet toy model, it is straightforward to see that it can be coupled to every SM operator at dimension five and, by comparing with Eq.\,\eqref{Eq:eDMEFT_Lagrangian}, we find 
\begin{equation}
    \kappa^\prime_\text{singlet} = \lambda_\kappa \frac{v}{\Lambda}, \quad (c^\prime_{q,\ell})_\text{singlet} = \lambda_{q,\ell} \frac{v}{\Lambda}, \quad (c_{X})_\text{singlet} = \lambda_X \frac{1}{\Lambda}, \quad (c_T^\prime)_\text{singlet} = \lambda_{c_T} \frac{v^3}{\Lambda^3} \, ,
    \label{eq:scalingSinglet}
\end{equation}
where $\lambda_i$ is expected to be an $\mathcal{O}(1)$ factor. We call this model (I). For a doublet and triplet the situation is more intricate, since contracting $(\phi^\dagger\phi)$ forms a singlet, but as long as $\langle\phi\rangle = 0$ no processes with a single physical field can be induced, as necessary for the resonant collider searches, see Appendix \ref{App:TableWilsonCoefficients}.  Therefore, we rely on invariant operators containing only one $\phi$.\footnote{It might be interesting to analyze signals stemming from the pair-production of scalars in a future work, should hints for such a scenario arise.} The simplest possibility is the Yukawa term, since an $SU(2)$ doublet can couple to the SM fermions without suppression, and we find 
\begin{equation}    \kappa^\prime_\text{doublet} = 0, \quad (c^\prime_{q,\ell})_\text{doublet} = \lambda_{q,\ell}, \quad (c_{X})_\text{doublet} = 0 , \quad (c_T^\prime)_\text{doublet} = 0 \, ,
    \label{eq:scalingDoublet}
\end{equation}
which we will call model (II). For a triplet the same Yukawa-type operator arises at dimension five by contracting the Higgs doublet with the LH fermion doublet ($\bf{2} \otimes \bf{2} = \bf{3} \oplus \bf{1}$) and contracting the resulting triplet with the new scalar triplet. The same argument applies for $\mathcal{O}_\kappa$ and $\mathcal{O}_{c_T}$ if the Higgs covariant derivative is contracted accordingly. Furthermore, it can couple to the $\mathcal{O}_{WB}$ operator in which $W_{\mu \nu} B^{\mu \nu}$ form a triplet. Then,
\begin{equation}
    \kappa^\prime_\text{trip} = \lambda_\kappa \frac{v}{\Lambda}, \ \ (c^\prime_{q,\ell})_\text{trip} = \lambda_{q,\ell} \frac{v}{\Lambda}, \ \ (c_{WB})_\text{trip} = \frac{\lambda_{WB}}{\Lambda}, \ \ (c^\text{rest}_{X})_\text{trip} = 0, \ \ (c_T^\prime)_\text{trip} = \lambda_{c_T}\frac{v^3}{\Lambda^3}  \, ,
    \label{eq:scalingTriplet}
\end{equation}
which will be denoted as model (III).  Interestingly, a triplet under these conditions behaves like a singlet unless a specific singlet UV-completion leads to a contribution from $c_X^\text{rest}$ that dominates over $(c_{WB})$. On average, we do not expect that and thus treat the triplet and singlet scaling the same in the following. To compare the three models with the best-fit obtained in Sec.\,\ref{sec:mod-indAnalysis} we define so-called ``Model Regions'', which is the parameter space we expect the models to live in if the $\mathcal{O}(1)$ parameters $\lambda_i$ are in fact $\mathcal{O}(1)$, e.g. in the interval [1/3,3]. 
\paragraph{Model Regions}
\begin{figure}[b!]
    \centering
    \begin{subfigure}{0.6\textwidth}
    \includegraphics[width=\linewidth]{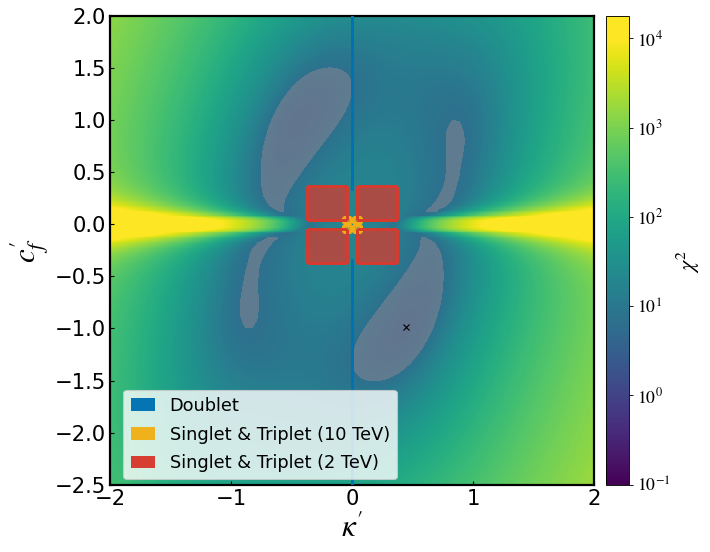}
    \caption{$c^\prime_f,\,\kappa^\prime$ and $\chi^2_{\text{bf}} = 2.1$}
    \label{fig:naturUniversal}
    \end{subfigure}
    \medskip
    \begin{subfigure}{0.49\textwidth}
    \includegraphics[width=\linewidth]{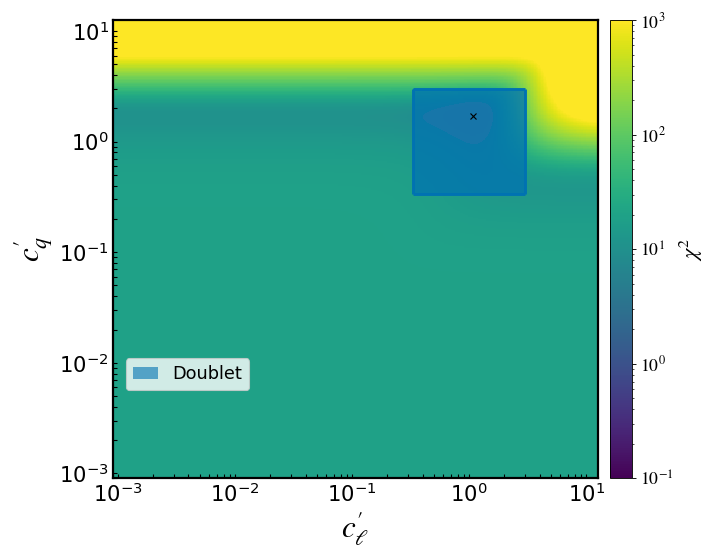}
    \caption{$\kappa^\prime=0,\,c^\prime_q,\,c^\prime_\ell$ and $\chi^2_{\text{bf}} = 4.2$}
    \label{fig:naturKappa0}
    \end{subfigure}
    \hfill
    \begin{subfigure}{0.49\textwidth}
    \includegraphics[width=\linewidth]{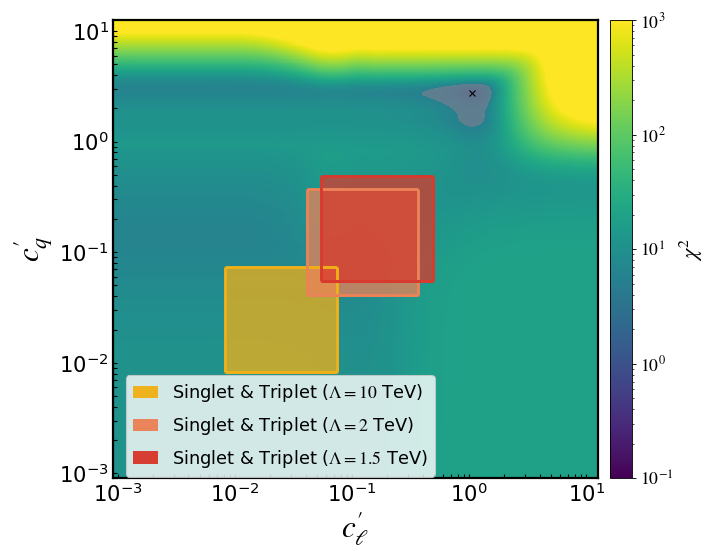}
    \caption{$\kappa^\prime \lesssim 0.5,\,c^\prime_q,\,c^\prime_\ell,\,c_X = 0.24$ TeV$^{-1}$ and $\chi^2_{\text{bf}} = 0.7$}
    \label{fig:natureMarginalize}
    \end{subfigure}
    \caption{Model regions compared to best-fit and $90\%$ C.L. region, see text for details.}
    \label{fig:naturalness}
\end{figure}
Fig.\,\ref{fig:naturalness} shows the comparison of the best-fit with the regions the model classes (I)-(III) are expected to live in.\footnote{Fig.\,\ref{fig:naturUniversal} corresponds to Fig.\,\ref{fig:chisquareUniversal}, Fig.\,\ref{fig:naturKappa0} to Fig.\,\ref{fig:chisquareKappazero} and Fig.\,\ref{fig:natureMarginalize} to Fig.\,\ref{fig:marginalization_small}.} The color-coding is now modified to correspond to the same $\chi^2$ values across the plots. The eHEFT analysis systematically shows  what the requirements for the coefficients are. For small couplings, the $\chi^2$ becomes independent of the precise values, i.e. the color in the plot remains the same. This is a remnant of the compatibility of the observed excesses with zero. Moreover, besides fitting the excess, when considering specific models one should clearly take into account additional constraints on the matched EFT coefficients stemming from further phenomenological bounds, e.g. null searches in other channels, and see if the excluded parameter space is in conflict with the model region. For Fig.\,\ref{fig:naturUniversal} a universal (aligned) fermion coupling is assumed and the coefficient related to the coupling to fermions is plotted against the coefficient related to the coupling to gauge bosons. The best fit has $\chi^2 = 2.1$ and although model (II) (in blue) predicts zero direct coupling to gauge bosons the model region overlaps with the 90\% C.L. region since it allows large couplings to fermions necessary to match the 95 GeV excesses. Models (I) and (III) (in red and yellow) predict similar coefficients for the couplings to fermions and gauge bosons, both generated at canonical dimension five. The larger $\Lambda$ for the singlet/triplet, the further the model is pushed from the 90\% C.L. region, as can be seen by comparing the red to the yellow model region, indicating that such an interpretation is viable already for $\Lambda = 2\,\mathrm{TeV}$ and becomes increasingly favored for lower new-physics scales, where the resonance effects are enhanced. 

Next, we allow for a non-universal fermion coupling and separate $c^\prime_f$ into a coupling to quarks $c^\prime_q$ and a coupling to leptons $c^\prime_\ell$. In Fig.\,\ref{fig:naturKappa0} $\kappa^\prime =0$ is fixed to align with the expectation for model (II), but $c_q^\prime$ and $c_\ell^\prime$ are varied individually, leading to a best-fit with $\chi^2_\mathrm{bf} = 4.2$. The trend observed before remains; a doublet model has no difficulties generating large fermion couplings required by the observed excesses. In Fig.\,\ref{fig:natureMarginalize} we marginalize over $\kappa^\prime$, but in a range bounded from above by $0.5$, the largest value consistent with the generation of the operator at canonical dimension five\footnote{The lowest $\Lambda$ considered here is 1.5 TeV and with $\lambda_i \in [1/3,3]$ we find $\kappa^\prime \lesssim 3 \times v/(1.5 \text{ TeV}) \sim 0.5$.} and $\Lambda \gtrsim 1.5$ TeV, set the coefficients for the chiral dimension four operators to $c_X = 0.24 \text{ TeV}^{-1}$,\footnote{This is compatible with the operators being generated at mass dimension five, with the lower end of the expected range with $\Lambda = 1.5$ TeV ($c_X = (1/3) /(1.5 \text{ TeV})$) and with upper end of the expected range with $\Lambda = 10$ TeV ($c_X = 3 /(10 \text{ TeV})$). We have further checked that even going towards maximal values of $c_X \sim 3/(1.5)$ TeV does not significantly change the best-fit results.} and find $\chi^2_{\text{bf}} = 0.7$, the lowest $\chi_\mathrm{bf}^2$ of the plots in Fig.\,\ref{fig:naturalness} but not overlapping with the model regions. 

Comparing the figures, we see that under the assumptions of our models, a doublet can match the best-fit naturally, although the $\chi^2$ value of the best-fit is comparably larger than when allowing for $\kappa^\prime \neq 0$. For both singlet and triplet there is a strong preference for a low NP scale due to the large fermion couplings in the best-fit regions. Additionally, different quark and lepton couplings lead to a better fit, i.e. lower value of $\chi^2$, than the strict flavor-alignment scenario considered in Fig.\,\ref{fig:naturUniversal}. However, realistic models have difficulties generating large enough fermion couplings to reach the regions of such low $\chi^2$. Instead, their naturalness regions coincide with $\chi^2$ values around $10^1$ at best, slightly worse than for the considered doublet. In the following section we will describe the properties an ideal model should have. We want to highlight that the best-fit was carried out only three times for three distinct scenarios (universal $c^\prime_f$ and $\kappa^\prime$, $\kappa^\prime = 0$ with non-universal $c^\prime_f$, marginalized $\kappa^\prime \lesssim 0.5$ with non-universal $c^\prime_f$), but that it is matched to significantly more model-incarnations; model~(I),~(II),~(III) with universal fermion coupling, model~(II) with non-universal fermion coupling, model~(III) with non-universal fermion coupling and different $\Lambda$, model~(I) with non-universal fermion coupling and different $\Lambda$. The list is non-exhaustive. For instance, for any model that predicts $\kappa=0$ and has at most subdominant effects from the $c_X$, the model regions could also be placed in Fig.\,\ref{fig:naturKappa0} and compared with the best-fit region. As long as the minimal assumptions going into the fit are met, for any UV-model one simply has to determine the behavior and range of the coefficients, keeping the model-specific constraints in mind, and overlay the expected and allowed parameter space regions onto the plot. If all parameters of Eq.~\eqref{Eq:eDMEFT_Lagrangian} would be included, even a single (multi-dimensional) fit would suffice. Our three fits would then correspond to three slices of the multidimentional fit.

\subsection{Hints Towards a Better Model}
\label{sec:betterModel}
Comparing the $\chi^2$ when restricting the EFT coefficients to lie in the region allowed by models (I)-(III), we see that it is in fact worse than the best-fit $\chi^2$ obtained in Sec.\,\ref{sec:mod-indAnalysis}. The reason is apparent; generally, a fairly large value of $\kappa^\prime \sim 0.6$ with $\chi^2 \sim 10^{-6}$ (or alternatively, an even larger $\kappa^\prime \sim 1.6$ with $\chi^2\sim10^{-2}$) is preferred to best match the 95 GeV excesses, as determined in Sec.\,\ref{sec:mod-indAnalysis}. For the singlet or triplet of models (I) and (III), $\kappa^\prime = \mathcal{O}(1) \, v/\Lambda$ which would imply either a tuned $\lambda > \mathcal{O}(1)$ or a low $\Lambda$, potentially problematic phenomenologically. For model (II), $\kappa^\prime = 0$. Therefore, which properties would a scalar have to possess to obtain $\kappa^\prime \sim 0.6$ while keeping a hierarchy $\Lambda \gg v$? Ideally, the covariant derivative operator should be marginal, and therefore the field no singlet under $SU(2)$. For the SM Higgs for instance, $\kappa^\prime = 2$.\footnote{The 2 originates from $(h+v)^2 \supset 2 v h$.} Then, furthermore keeping in mind that the term in Eq.\,\eqref{Eq:eDMEFT_Lagrangian} is normalized to the Higgs vev, to find the relevant coefficient we have to replace the insertion of the Higgs vev by a new vev, in which case we expect
\begin{equation}
    \kappa^\prime \propto \frac{v_{\text{new}}}{v}.
\end{equation}
Then, $\kappa^\prime \sim 0.6$ can be achieved by an according ratio between the vev of the new field $v_{\text{new}}$ and the Higgs vev. Specifically, for different multiplets $\kappa^\prime$ behaves as
\begin{equation}
    (\kappa^\prime)^{\text{doublet}} \sim \frac{2 v_{\text{new}}}{v}, \quad  (\kappa^\prime)^{\text{trip}} \sim \frac{4 \, v_{\text{new}}}{v}, \quad ...
\end{equation}
so a triplet seems more favorable to realize large $\kappa^\prime$ than a doublet, see e.g. \cite{Ashanujjaman:2023etj} for an $SU(2)$ triplet explanation of the 95 GeV excess. One might think that going to even larger multiplets could be a way forward, but this would be at the cost of suppressing the fermion couplings, since they have to be generated at canonical dimension six as $\bar{L}_L H q_R (S^\dagger S/\Lambda^2) \Rightarrow c^\prime_f \sim (v v_S / \Lambda^2)$. If combined with a low $\Lambda$, requiring a large $v_S$ might even move the EFT out of the linear limit, in which case additional operators could become relevant. Additionally, one has to be careful to not induce a contribution to the W boson mass that is in conflict with observations, leading to strong limits on possible values of $\kappa^\prime$ for new scalars that obtain a vev. It would be interesting to build a consistent model where the interaction strength approaches $\kappa^\prime \sim 0.6$ while avoiding a phenomenologically problematic W-boson mass.\footnote{There is a small tension between the result obtained by CDFII \cite{CDF:2022hxs} and the recent ATLAS\cite{ATLAS:2023fsi} and LHCb \cite{LHCb:2021bjt} results, but even generously enlarging the error, as is done in \cite{Crivellin:2023xbu}, and comparing to the SM prediction there is very little room to realize a comparatively large NP contribution to the W-boson mass.} Another possible avenue could be to boost the chiral dimension-four operators and therefore allow for a smaller contribution of $\kappa^\prime$ to the di-photon resonance, although we have checked that for realistic values of $c_X$ (Fig.\,\ref{fig:marginalization_small}) there is still a preference to maximize $\kappa^\prime$.\footnote{The effect that including the chiral dimension four operators has is shown in Appendix \ref{app:operatorsChDim4} over a wide mass range.} Furthermore, we have seen that $\chi^2_{\text{bf}}$ is lower for $c^\prime_\ell \neq c^\prime_q$, hinting towards a model with a slightly enhanced coupling to quarks compared to leptons. Generally, large fermion couplings are preferred, thus pointing towards an $SU(2)$ doublet for which a Yukawa coupling can arise at mass dimension four. 

Since the 95 GeV resonance is used only as a toy example, in order to showcase how this analysis framework can be employed in the future, here we do not delve further into working out the details of this better model. Still, it is interesting to see how this analysis naively points towards a field charged under $SU(2)$ with a fairly large non-zero expectation value, or alternatively towards a singlet with either a low $\Lambda$ or a coefficient $\lambda_i$ larger than ${\cal O}(1)$. Of course, there is the possibility that no consistent scalar model with these properties exists, i.e. without being in conflict with other phenomenological constraints, perhaps hinting at the fact that the signal was caused by a pseudoscalar instead \cite{Iguro:2022dok}, that the eHEFT should be considered in the non-linear limit, or that one or more of the observed excesses might have been statistical fluctuations rather than a signature of a new field \cite{Aguilar-Saavedra:2023tql}. The eHEFT analysis can be an important tool to judge and understand fundamental properties of any collider excess observed in the future. 

\section{Outlook to Other Mass Ranges}
\label{sec:outlook}
In the moment in which a signal is found, the channels which have to be fitted are determined by the observation of excesses. The procedure then follows the $95$ GeV example in Sec.\,\ref{sec:95GeVres}. In the meantime, we display three specific signatures for the outlook to other mass ranges, which we expect to be particularly promising in order to understand the structural relations between couplings to gauge bosons and to fermions and therefore the $SU(2)$ structure of the new field. They are $\tau$-pair production, di-boson and di-photon resonances. The expected behavior for other decay channels can easily be derived using the decay widths given in Appendix \ref{app:decayWidths}. The decay into photons, although not necessarily dominant, is a very clear signal experimentally, and it often proceeds via fermion loops, predominantly top quark loops, as well as $W$-boson loops, making it the ideal candidate to probe the interplay of those couplings. Furthermore, since there is no leading-order contribution, the effect of higher dimensional operators can be comparable to the SM loop-induced process thus making the decay into two photons a sensitive probe of the underlying structure. As an example for a scalar mass around 650 GeV and SM-like couplings the top and $W$-loops destructively interfere, leading to a dip in the decay width, as is well known. Instead, a scalar with additional terms contributing to the decay, or a different hierarchy of couplings, can shift or even lift the degeneracy completely. Then, to probe the fermion coupling individually we have chosen to focus on the decay into $\tau\bar{\tau}$, although for frameworks following SM Yukawa hierarchies we would expect a larger coupling to quarks than leptons, since hadronic decays are notoriously messy. Lastly, for the di-boson resonance only the branching ratio into $W$ bosons is shown, to keep the plots as essential as possible. 

\subsection{Expected Behavior -- CP-even}
In the following, we focus on the three models detailed in Sec.\,\ref{sec:specificModels}, and scan the branching ratios into taus, W-bosons and photons. By plotting the branching ratios almost all dependence on $\Lambda$ drops out, the reason being that in the considered models most non-zero parameters carry the same power of $\Lambda$, apart from $c_T^\prime$ which arises at higher mass dimension when generated and for which we have checked that it has negligible effects on the branching ratios, see Appendix \ref{app:decayZZ}. This is not generally true for any model. For the scan we randomly generate $10^5$ points with mass between [100,1000] GeV, and $\mathcal{O}(1)$ parameters $\lambda_i$ between $\pm[1/3,3]$. Additionally, the lines for $\lambda_i = 1$ are shown. 

\paragraph{Model (I)}
\begin{figure}
    \centering
    \begin{subfigure}{0.49\textwidth}
    \includegraphics[width=\linewidth]{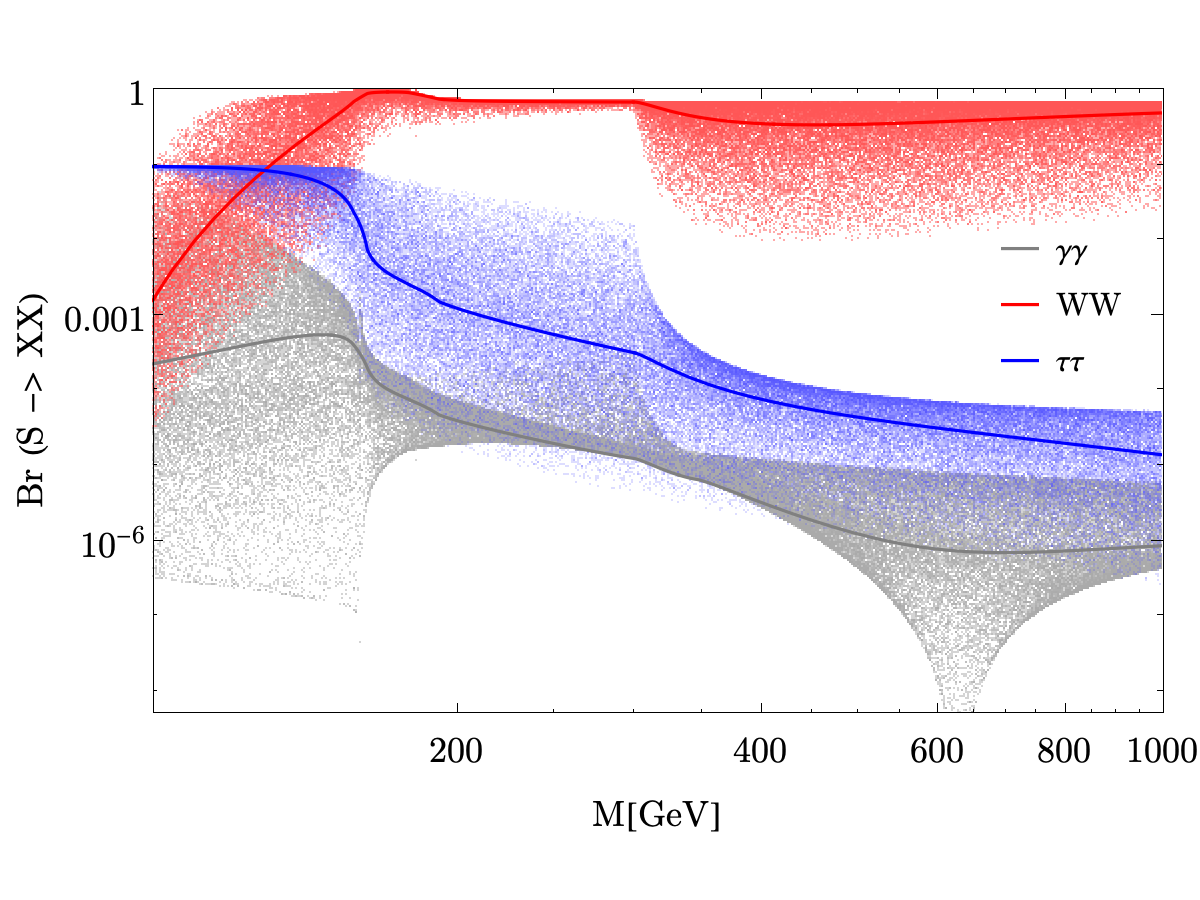}
    \caption{$\lambda_X, \lambda_{c_T} = 0$}
    \label{fig:BrSinglet}
    \end{subfigure}
    \hfill
    \begin{subfigure}{0.49\textwidth}
    \includegraphics[width=\linewidth]{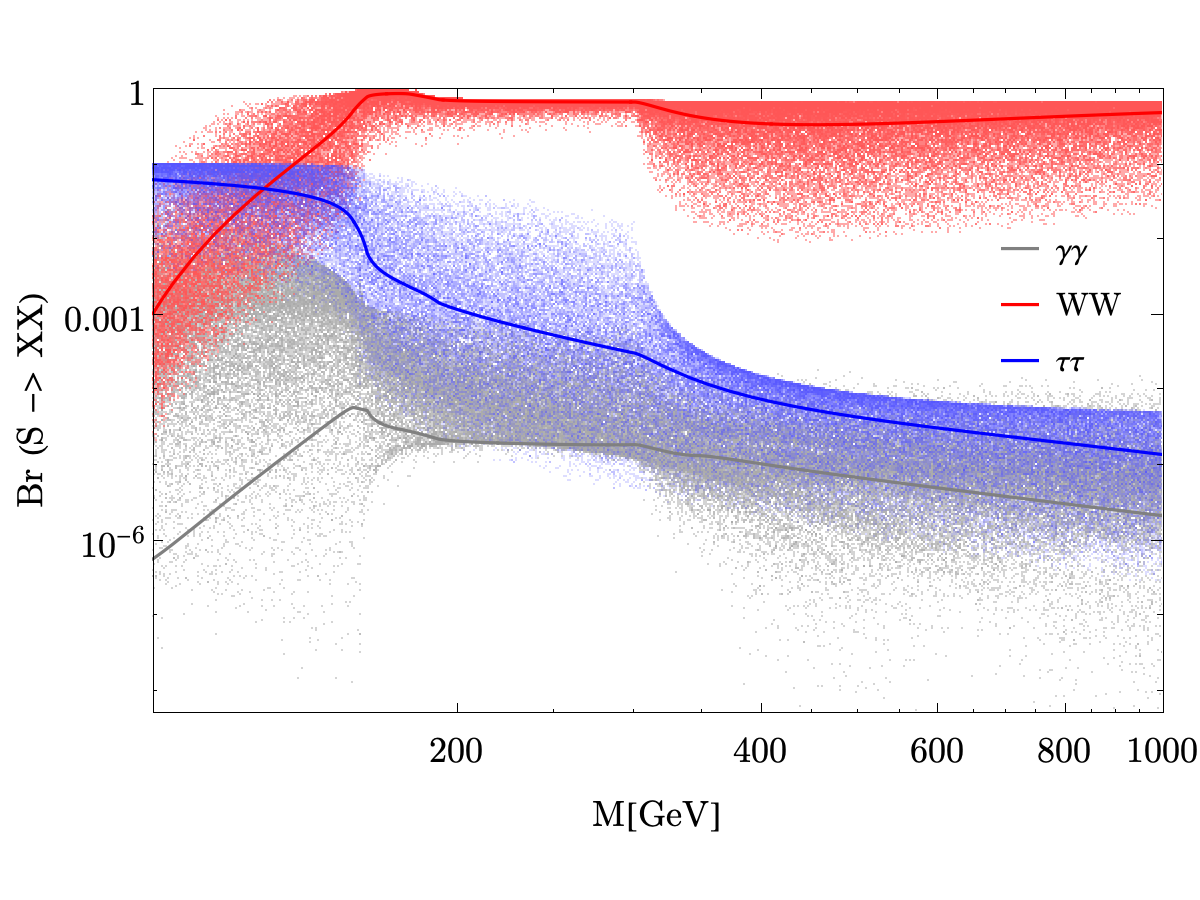}
    \caption{$\lambda_X, \lambda_{c_T} \sim \mathcal{O}(1)$}
    \label{fig:BrSingletFST}
    \end{subfigure}
    \caption{Expected branching ratios for model (I), varying the $\mathcal{O}(1)$ factors $\lambda_i$ of the eHEFT between $\pm$[1/3,3]. The lines show the prediction for $\lambda_i =1$. Fig.\,\ref{fig:BrSinglet} shows the leading order prediction, see Appendix \ref{app:decayZZ} for the effect of the chiral dimension two operator $\mathcal{O}_{c_T}$, while Fig.\,\ref{fig:BrSingletFST} includes all operators of Eq.\,\eqref{Eq:eDMEFT_Lagrangian}. The fermion coupling is taken to be universal and flavor aligned.}
    \label{fig:singletboth}
\end{figure}
The branching ratios of the singlet receive contributions from the $\mathcal{O}(1)$ parameters $\lambda_f,\lambda_\kappa, \lambda_X, \lambda_{c_T}$, see Eq.\,\eqref{eq:scalingSinglet}. Fig.\,\ref{fig:BrSinglet} shows the expected behavior without taking into account operators of chiral dimension four and the (suppressed) ${\cal O}_{c_T}$, therefore setting $\lambda_X = \lambda_{c_T} = 0$, while Fig.\,\ref{fig:BrSingletFST} shows the expected behavior when including all terms present in the Lagrangian in Eq.\,\eqref{Eq:eDMEFT_Lagrangian}. Without the additional coupling to the field strength tensors, the singlet couplings mimic the SM Higgs couplings, apart from being suppressed by $\Lambda$, which, as laid out above, drops out when considering branching ratios.\footnote{The overall cross sections are sensitive to the suppression.}  It is therefore no surprise to see that for scalar masses around 650 GeV, the decay into photons can be suppressed. A partial cancellation can also occur for $m_S \lesssim 2 \, m_W$ when $\lambda_\kappa \sim 0.4 \lambda_f$. However, once additional operators contribute to the decay, the situation changes. The couplings to the field strength tensors generate a coupling to photons that is expected to be sizable enough to counteract the destructive interference, even though the according coefficient sees an additional loop suppression. Still, since the decay into photons that is induced by the chiral dimension four operators is proportional to $(\lambda_W + \lambda_B - \lambda_{BW})$, a partial cancellation can be restored which is why a few of the parameter points predict very low $S \rightarrow \gamma \gamma$ in the high mass range.\footnote{We show the effect of $\lambda_{c_T}$ in Appendix \ref{app:decayZZ}.} Therefore, even if stringent limits on the decay into photons would be found experimentally, this unfortunately cannot exclude the generation of the operators in the last line of the Lagrangian in Eq.\,\eqref{Eq:eDMEFT_Lagrangian}. 

\begin{figure}
    \centering
    \begin{subfigure}{0.49\textwidth}
    \includegraphics[width=\linewidth]{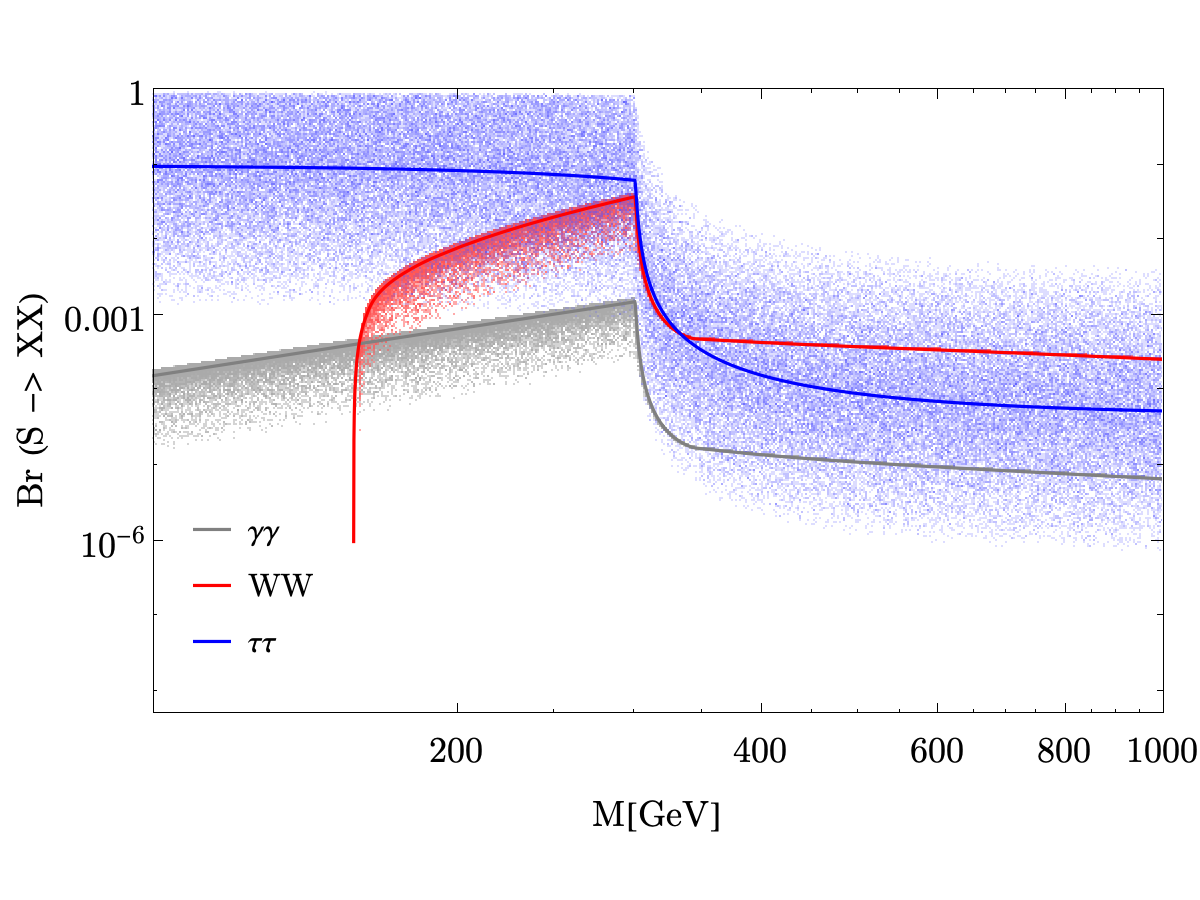}
    \caption{Doublet}
    \label{fig:BrDoublet}
    \end{subfigure}
    \hfill
    \begin{subfigure}{0.49\textwidth}
    \includegraphics[width=\linewidth]{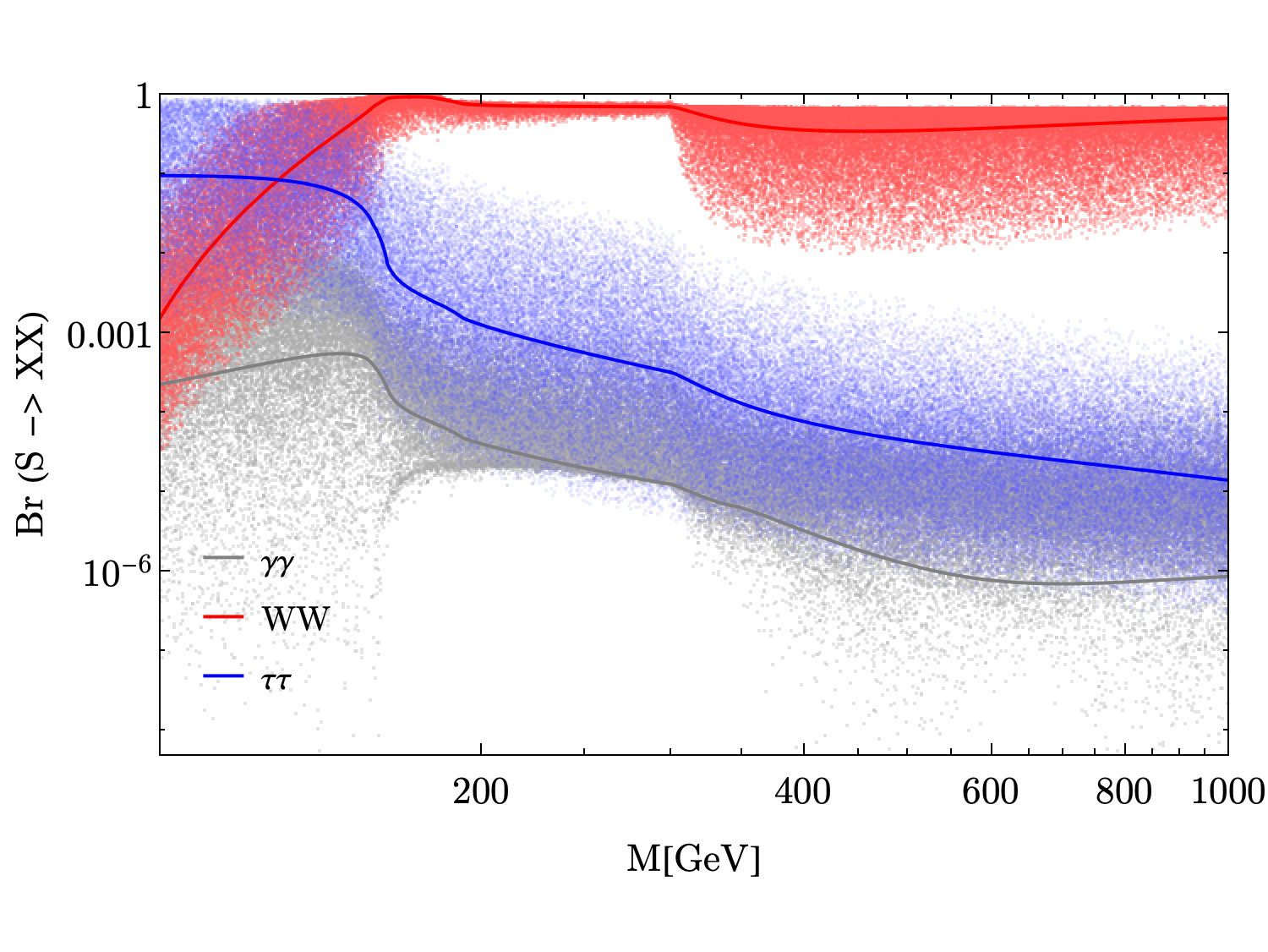}
    \caption{Triplet}
    \label{fig:BrTriplet}
    \end{subfigure}
    \caption{Expected branching ratios for models (II) and (III), varying the $\mathcal{O}(1)$ factors $\lambda_i$ of the eHEFT between $\pm$[1/3,3]. The lines show the prediction for universal $\lambda_i = 1, \, \lambda^\text{trip}_{WB} = 0$.}
    \label{fig:BrDoubletTriplet}
\end{figure}

\paragraph{Model (II)}
The branching ratios of the doublet receive contributions only from the $\mathcal{O}(1)$ parameter $\lambda_f$, see Eqs.\eqref{eq:scalingDoublet}. We allow for different couplings to quarks versus leptons, and scan over $\lambda_q$ and $\lambda_l$. A universal $\lambda_f$ would be a global factor for these types of doublets (remember that the $WW$ decay is induced via a fermion loop), which the branching ratios are not sensitive to, and varying it produces points laying exactly on the lines shown in the Fig.\,\ref{fig:BrDoublet}. If such a scenario were realized, we could make very precise predictions for the ratio between branching ratios to compare to LHC excesses.
\paragraph{Model (III)}
The branching ratios of the triplet are very similar to the singlet case, as explained above. In Fig.\,\ref{fig:BrTriplet} we show the predictions, and see only slight differences, related to the missing contributions from $\mathcal{O}_{WW},\,\mathcal{O}_{BB}$, in particular in the di-photon channel. In a particular UV-completion, the scaling of one of the models might change, e.g. by not generating a $\kappa^\prime_\text{trip},$ which would allow to better differentiate the models.

\subsubsection{General Predictions}
Comparing Fig.\,\ref{fig:singletboth} and Fig.\,\ref{fig:BrDoubletTriplet}, there are certain trends that can be observed. For instance, if a low mass scalar is found where the decay into gauge bosons dominates over the decay into taus, this does not seem to be well described by model (II). In contrast, for a high mass scalar when the decay into gauge bosons is subdominant with respect to the decay into taus or photons, models (I) or (III) would not be a good match. The here-presented outlook is simplified; once mixing with the Higgs is no longer neglected the predictions are washed out due to additional parameters. Still, it is interesting to look at the general behavior to qualitatively understand how the $SU(2)$ nature influences branching ratios.  
Indeed, when looking at the ratio between branching ratios a lot of systematic uncertainties are expected to drop out, for instance related to the production of the scalar, making the ratio a promising and sensitive probe of the nature of the new scalar field. It should be mentioned that the framework loses its predictive power if only a single excess is observed, since it becomes hard to disentangle the suppression stemming from the scale of $\Lambda$ and suppression due to the canonical dimension at which the relevant operator is generated. For qualitative statements it is essential to have a handle on the ratio between the coefficients. 

\subsection{Expected Behavior -- CP-odd}
If the new scalar is CP-odd, the decay widths are modified, leading to slightly different predictions for the branching ratios. The coefficients in models (I)$_{\text{ps}}$, (II)$_\text{ps}$ and (III)$_\text{ps}$ for the pseudo-scalar case are
\begin{equation}
    \kappa^\prime_\text{(I)$_{\text{ps}}$} = 0, \quad (c^\prime_{q,\ell})_\text{(I)$_{\text{ps}}$} = \lambda_{q,\ell} \frac{v}{\Lambda}, \quad (\tilde{c}_{X})_\text{(I)$_{\text{ps}}$} = \lambda_X \frac{1}{\Lambda}, \quad (c_T^\prime)_\text{(I)$_{\text{ps}}$} = 0 \, ,
    \label{eq:scalingSingletps}
\end{equation}
\begin{equation}
    \kappa^\prime_\text{(II)$_{\text{ps}}$} = 0, \quad (c^\prime_{q,\ell})_\text{(II)$_{\text{ps}}$} = \lambda_{q,\ell}, \quad (\tilde{c}_{X})_\text{(II)$_{\text{ps}}$} = 0, \quad (c_T^\prime)_\text{(II)$_{\text{ps}}$} = 0 \, ,
    \label{eq:scalingDoubletps}
\end{equation}
\begin{equation}
    \kappa^\prime_\text{(III)$_{\text{ps}}$} = 0, \quad (c^\prime_{q,\ell})_\text{(III)$_{\text{ps}}$} = \lambda_{q,\ell} \frac{v}{\Lambda}, \quad (\tilde{c}_{WB})_\text{(III)$_{\text{ps}}$} = \frac{\lambda_{WB}}{\Lambda}, \quad (\tilde{c}^\text{rest}_{X})_\text{(III)$_{\text{ps}}$} = 0, \quad (c_T^\prime)_\text{(III)$_{\text{ps}}$} = 0 \, .
    \label{eq:scalingTripletps}
\end{equation}
As before, a triplet behaves almost like a singlet. By requiring CP-invariance $\kappa^\prime, c_T^\prime = 0$ independently of $SU(2)$ nature. Note, however, that it would be possible to build a canonical dimension six operator by coupling $S^\dagger S / \Lambda^2$ to the Higgs covariant derivative, although not relevant for the phenomenology studied here. 

Comparing the scaling of the pseudoscalar models (I)$_\text{ps}$ - (III)$_\text{ps}$ (Eqs.\eqref{eq:scalingSingletps}, \eqref{eq:scalingDoubletps}, \eqref{eq:scalingTripletps}) to the scaling of the scalar models (I) - (III) (Eqs.\eqref{eq:scalingSinglet}, \eqref{eq:scalingDoublet}, \eqref{eq:scalingTriplet}), there is one difference: a CP-even singlet can couple to the Higgs covariant derivatives at canonical dimension five, while any CP-odd multiplet needs to be added as $(S^\dagger S)$ to respect CP-invariance, thus not generating a trilinear coupling relevant for the here-considered phenomenology. In general, observing a dominant decay into electroweak gauge bosons points towards a CP-even scalar. For the CP-odd case the different $SU(2)$ representations behave similarly, in that the decay to fermions will be dominant. Note that additional phenomenologically relevant contributions to the $VVS$ gauge couplings of a CP-odd scalar may arise for the singlet and triplet representations through higher-dimensional operators, such as $i \tilde{B}_{\mu \nu} \operatorname{Tr}\left(\Sigma^\dagger D^\mu\Sigma \sigma^3\right) \partial^\nu \mathcal{F}(S)$ and $i \operatorname{Tr}\left(\tilde{W}_{\mu \nu} D^\mu\big(\Sigma\big) \Sigma^\dagger\right) \partial^\nu \mathcal{F}(S)$. Although suppressed as the coefficient $c_X$, these operators would induce an additional qualitative difference with respect to the doublet case. In the moment in which a signal has to be matched, and the field produced at a collider, there will be a preference for a particular value for the coefficients which in turn is related to the canonical dimension of the corresponding operators, i.e. to the $SU(2)$ representation the scalar originates from. To differentiate a pseudoscalar from a scalar resonance, it is useful to consider more than just signal strengths. For instance, the angular distribution of an event can give insights into the CP structure, see e.g. \cite{Dolan:2014upa}. Furthermore, as explained in Sec.\,\ref{sec:Framework}, the derivative operator in the second line of Eq.\,\eqref{Eq:eDMEFT_Lagrangian} can generate a $Z h S$ interaction only for a pseudoscalar \cite{Bauer:2016zfj}. 

\section{Conclusion}
\label{sec:conclusion}
We have presented the extended Higgs Effective Field Theory (eHEFT) framework, which includes a new collider-accessible scalar degree of freedom at low energies in addition to the SM content. By implementing the EW symmetry non-linearly, we allow for the possibility that the new scalar originated from a non-trivial $SU(2)$ multiplet where heavier modes have been integrated out, all while keeping the description gauge-invariant. This makes the eHEFT well suited for a general (and consistent) analysis of collider signatures. Depending on the UV properties of a model, specifically the $SU(2)$ nature of the scalar, the eHEFT operators are generated at distinct canonical mass dimension.  Then, when a collider anomaly is found, the signal strength can be fitted within the eHEFT and the resulting best-fit values and hierarchies between the coefficients give valuable insights into the properties a complete model must have to explain the observed excess. The procedure works in two directions; one can derive the required behavior of the eHEFT coefficients and infer the required UV properties, or one can start with a specific model in mind, match it to the eHEFT and see whether it is consistent with the parameter space obtained during the fitting. Matching to the eHEFT is more efficient than carrying out a statistical analysis for each potential model individually, and several model incarnations can be matched to the same eHEFT fit, thus facilitating the comparison of different models. 

The eHEFT discussed here includes effects up to NLO, i.e. operators at chiral dimension two and resulting loops, as well as tree-level contributions from chiral dimension four operators, for which we included only phenomenologically relevant operators. We assume new physics apart from $S$ to lie beyond the collider reach and thus expect operators to experience a suppression when arising at higher mass dimension, see the discussion in the power-counting paragraph of Sec.\,\ref{sec:Framework}. If the new field were to obtain a large vev in combination with a low scale of NP for instance, this assumption might break down. In this case further chiral-dimension four operators should be added to the Lagrangian before proceeding with the eHEFT method. However, for a large set of models this is not the case, and they can already be well-described by the truncated Lagrangian in Eq.\,\eqref{Eq:eDMEFT_Lagrangian}. The Lagrangian is formulated to respect flavor alignment, but by rescaling the fermionic eHEFT coefficients this assumption can be easily modified. Then, the analysis goes as follows; first, one or more collider signals have to be chosen for the fitting, next, the best-fit is determined by scanning over the dimensionless eHEFT coefficients in a suitable range which leads to a parameter space for the coefficients that matches the observed excess. From this, one can either infer what is required by a UV-model or check whether a specific UV-model can lie in that parameter space without being in conflict with other phenomenological constraints. 

As an illustrative example, the eHEFT analysis has been carried out for the tentative 95 GeV resonance in Sec.\,\ref{sec:95GeVres}, showing that there is preference for the model to have non-universal fermion couplings, and although the signals are compatible (at 90\% C.L.) with $\kappa^\prime = 0$, i.e. no direct coupling to the gauge bosons, the lowest $\chi^2$ value was obtained for $\kappa^\prime \sim 0.6$, comparable to the SM value of $\kappa_{\text{SM}}^\prime = 2$. Next, we showed how to compare models to the fitted coefficients, using three types of models (I)-(III). To illustrate the relation between the $SU(2)$ nature and expected coefficients, we chose simple toy models. Then, the natural parameter space for each model is determined by an $\mathcal{O}(1)$ number and the canonical dimension at which an operator can be generated. We considered a singlet, doublet and triplet under $SU(2)$. By varying the $\mathcal{O}(1)$ parameter we find so-called ``model" regions, which showcase how well the model can describe the LHC excess. Depending on the suppression scale $\Lambda$, the model regions move in parameter space and can approach or distance themselves from the best-fit regions. For more realistic models, the same will be true, albeit with a few more free parameters. In 2HDM models for instance, the angles $\alpha, \beta$ will have an effect on where the model region sits. Consequently, the model regions can give a qualitative understanding of how specific parameters influence the ability of a model to recreate the signal, and combining them with the best-fit regions furthermore gives a quantitative preference for parameter values. For the models (I) - (III) we find that a doublet can quite naturally describe the observed 95 GeV excesses while a singlet and triplet would require very low $\Lambda$  given the assumptions detailed in Sec.~\ref{sec:specificModels}. However, the lowest value for $\chi^2$ was obtained outside of the parameter space allowed in the three models, since none can generate an unsuppressed coupling to gauge bosons. We sketched ways towards a better model in Sec.\,\ref{sec:betterModel}. In the future, it would be interesting to work out if consistent models can be built that results in large $\kappa^\prime$ without being in conflict with electroweak observables. 

The last part of the paper is dedicated to an outlook to other mass ranges. We have shown that we expect qualitatively very different behaviors for an $SU(2)$ singlet or triplet compared to a doublet, although it should be stressed once more that more complicated models with additional free parameters, such as mixing angles, can potentially wash out this effect. The eHEFT method is particularly predictive when two or more excesses are observed, since many uncertainties drop out when looking at a ratio of signals, leaving it a sensitive probe to the order at which operators are generated, that is to the $SU(2)$ nature of the scalar. 

In conclusion, the here-described framework is consistent, economical and can be used as a guide for interpreting LHC excesses in a least biased way. Extended scalar sectors are a motivated addition to the SM and with hints for new scalars continuously popping up,\footnote{In addition to the 95 GeV resonance, there are indications for a 650 GeV resonance, but since some of the excesses are in combination with the 95 GeV resonance \cite{CMS:2023boe,Ellwanger:2023zjc} in order to fully describe it the eHEFT has to be extended to include two additional scalar degrees of freedom, similar to how it was done in \cite{Arcadi:2024mli}. Furthermore, there is a strongly debated excess around 690 GeV \cite{Cline:2025umw,Consoli:2025nsv} and an emerging excess around $\sim150$ GeV \cite{Crivellin:2021ubm,Bhattacharya:2025rfr}, see for an overview e.g. Sec.\,(3.10) in \cite{Crivellin:2023zui}.} 
there is hope that one of them will solidify, requiring a structural way of interpreting the signal such as by using the eHEFT method. In the future, it would be interesting to extend the framework to allow for more than one collider-accessible new field, so that an even larger class of UV-models can be accurately described through the eHEFT, and to furthermore apply it to low-energy observables.

\section*{Acknowledgements}
We thank Sreejit Das, Sven Fabian, Pol Morell and Jo\~{a}o Paulo Pinheiro for useful discussions.
DCA acknowledges funding from the Spanish MCIN/AEI/10.13039/501100011033 through grant PID2022-136224NB-C21.

\clearpage
\begin{appendix}

\section{Wilson Coefficients for Collider Searches and Constraints} \label{App:TableWilsonCoefficients}
\begin{table}[h!]
    \centering
    \resizebox{\textwidth}{!}{
    \begin{tabular}{cccc}
        \toprule
        \multirow{2}*{Category}  & \multirow{2}*{Process} & \multicolumn{2}{c}{Model parameter}\\ & &Left vertex & Right vertex \\     \midrule
        \multirow{7}*{$\tau$ pair production }&\begin{tikzpicture}[baseline=12mm]
                    \begin{feynman}
                        \vertex (a1) {\(g\)};
                        \vertex[above=1.5cm of a1] (a2) {\(g\)};

                        \vertex at ($(a1)!0.5!(a2) + (1cm,0)$) (b1) [dot] {};
                        \vertex[right=1cm of b1] (c1);

                        \vertex[right=3cm of a1] (d1) {\(\tau^{+}\)};
                        \vertex[right=3cm of a2] (d2) {\(\tau^{-}\)};

                        \diagram* {
                        (a1) -- [gluon] (b1),
                        (a2) -- [gluon] (b1),
                        (b1) -- [scalar, edge label=\( \mathcal{S}\)] (c1),
                        (d1) -- [fermion] (c1),
                        (d2) -- [anti fermion] (c1)
                    };
                    \end{feynman}
                \end{tikzpicture}& \multirow{3}*{\makecell{ $c_{G}$
                }}
                & \multirow{3}*{\makecell{$c^\prime_\ell$}}\\
                & \begin{tikzpicture}[baseline=12mm]
                    \begin{feynman}
                        \vertex (a1) ;
                        \vertex[above=1.5cm of a1] (a2);

                        \vertex at ($(a1)!0.5!(a2) + (0.7cm,0)$) (b1) [dot] {};
                        \vertex[right=1cm of b1] (c1);

                        \vertex[left=1cm of a1] (y1){\(g\)};
                        \vertex[left=1cm of a2] (y2) {\(g\)};

                        \vertex[right=2.5cm of a1] (d1) {\(\tau^{+}\)};
                        \vertex[right=2.5cm of a2] (d2) {\(\tau^{-}\)};

                        \diagram* {
                        (y1) -- [gluon] (a1),
                        (y2) -- [gluon] (a2),
                        (a1) -- [fermion] (b1),
                        (a2) -- [fermion] (b1),
                        (a1) -- [fermion, edge label=\( q\)] (a2),
                        (b1) -- [scalar, edge label=\( \mathcal{S}\)] (c1),
                        (d1) -- [fermion] (c1),
                        (d2) -- [anti fermion] (c1)
                    };
                    \end{feynman}
                \end{tikzpicture}&\multirow{3}*{\makecell{$c^\prime_{q}$}}&\multirow{3}*{\makecell{$c^\prime_\ell$}} \\
                 \midrule
        
                \multirow{9}*{\makecell{Di-boson\\resonance}}& \begin{tikzpicture}[baseline=12mm]
                    \begin{feynman}
                        \vertex (a1) {\(g\)};
                        \vertex[above=1.5cm of a1] (a2) {\(g\)};

                        \vertex at ($(a1)!0.5!(a2) + (1cm,0)$) (b1) [dot] {};
                        \vertex[right=1cm of b1] (c1);

                        \vertex[right=3cm of a1] (d1) {\(Z/W^{-}\)};
                        \vertex[right=3cm of a2] (d2) {\(Z/W^{+}\)};

                        \diagram* {
                        (a1) -- [gluon] (b1),
                        (a2) -- [gluon] (b1),
                        (b1) -- [scalar, edge label=\(\mathcal{S}\)] (c1),
                        (d1) -- [boson] (c1),
                        (d2) -- [boson] (c1)
                    };
                    \end{feynman}
                \end{tikzpicture} & \multirow{3}*{\makecell{$c_G, c^\prime_{q}$}}& \multirow{3}*{\makecell{$c_W,c_B,c_{WB},\kappa^\prime,c_T^\prime$}}\\
                & \begin{tikzpicture}[baseline=12mm]
                    \begin{feynman}
                        \vertex (a1) {\(g\)};
                        \vertex[above=1.5cm of a1] (a2) {\(g\)};

                        \vertex at ($(a1)!0.5!(a2) + (1cm,0)$) (b1) [dot] {};
                        \vertex[right=1cm of b1] (c1);

                        \vertex[right=3cm of a1] (d1) ;
                        \vertex[right=3cm of a2] (d2);

                        \vertex[right=0.7cm of d1] (z1) {\(\ Z / W^+ \)};
                        \vertex[right=0.7cm of d2] (z2) {\( Z / W^-\)};

                        \diagram* {
                        (a1) -- [gluon] (b1),
                        (a2) -- [gluon] (b1),
                        (b1) -- [scalar, edge label=\(\mathcal{S}\)] (c1),
                        (d1) -- [fermion] (c1),
                        (d2) -- [fermion] (c1),
                        (d1) -- [boson] (z1),
                        (d2) -- [boson] (z2),
                        (d1) -- [fermion, edge label=\(\ f \)] (d2)
                    };
                    \end{feynman}
                \end{tikzpicture}&\multirow{3}*{\makecell{$c_G, c^\prime_{q}$}}&\multirow{3}*{\makecell{$c^\prime_q,c^\prime_\ell$}}\\
                & + $q$-loop production  &\multirow{3}*{\makecell{}}&\multirow{3}*{\makecell{}} \\
                \midrule
                \multirow{13}*{$\gamma \gamma$ production }&\begin{tikzpicture}[baseline=12mm]
                    \begin{feynman}
                        \vertex (a1) {\(g\)};
                        \vertex[above=1.5cm of a1] (a2) {\(g\)};

                        \vertex at ($(a1)!0.5!(a2) + (1cm,0)$) (b1) [dot] {};
                        \vertex[right=1cm of b1] (c1) [dot] {};

                        \vertex[right=3cm of a1] (d1) {\(\gamma\)};
                        \vertex[right=3cm of a2] (d2) {\(\gamma\)};

                        \diagram* {
                        (a1) -- [gluon] (b1),
                        (a2) -- [gluon] (b1),
                        (b1) -- [scalar, edge label=\(\mathcal{S}\)] (c1),
                        (d1) -- [boson] (c1),
                        (d2) -- [boson] (c1)
                    };
                    \end{feynman}
                \end{tikzpicture}& \multirow{3}*{\makecell{ $c_{G}, c^\prime_q$
                }}
                & \multirow{3}*{\makecell{$c_W, c_B,c_{WB}$}}\\
                & \begin{tikzpicture}[baseline=12mm]
                    \begin{feynman}
                        \vertex (a1) {\(g\)};
                        \vertex[above=1.5cm of a1] (a2) {\(g\)};

                        \vertex at ($(a1)!0.5!(a2) + (1cm,0)$) (b1) [dot] {};
                        \vertex[right=1cm of b1] (c1);

                        \vertex[right=3cm of a1] (d1) ;
                        \vertex[right=3cm of a2] (d2);

                        \vertex[right=0.7cm of d1] (z1) {\(\gamma\)};
                        \vertex[right=0.7cm of d2] (z2) {\(\gamma\)};

                        \diagram* {
                        (a1) -- [gluon] (b1),
                        (a2) -- [gluon] (b1),
                        (b1) -- [scalar, edge label=\(\mathcal{S}\)] (c1),
                        (d1) -- [boson] (c1),
                        (d2) -- [boson] (c1),
                        (d1) -- [boson] (z1),
                        (d2) -- [boson] (z2),
                        (d1) -- [boson,, edge label=\(\ W\)] (d2)
                    };
                    \end{feynman}
                \end{tikzpicture}&\multirow{3}*{\makecell{$c_G, c^\prime_{q}$}}&\multirow{3}*{\makecell{$\kappa^\prime$}} \\
                & \begin{tikzpicture}[baseline=12mm]
                    \begin{feynman}
                        \vertex (a1) {\(g\)};
                        \vertex[above=1.5cm of a1] (a2) {\(g\)};

                        \vertex at ($(a1)!0.5!(a2) + (1cm,0)$) (b1) [dot] {};
                        \vertex[right=1cm of b1] (c1);

                        \vertex[right=3cm of a1] (d1) ;
                        \vertex[right=3cm of a2] (d2);

                        \vertex[right=0.7cm of d1] (z1) {\(\gamma\)};
                        \vertex[right=0.7cm of d2] (z2) {\(\gamma\)};

                        \diagram* {
                        (a1) -- [gluon] (b1),
                        (a2) -- [gluon] (b1),
                        (b1) -- [scalar, edge label=\(\mathcal{S}\)] (c1),
                        (d1) -- [fermion] (c1),
                        (d2) -- [fermion] (c1),
                        (d1) -- [boson] (z1),
                        (d2) -- [boson] (z2),
                        (d1) -- [fermion, edge label=\(\ f\)] (d2)
                    };
                    \end{feynman}
                \end{tikzpicture}&\multirow{3}*{\makecell{$c_G, c^\prime_{q}$}}&\multirow{3}*{\makecell{$c^\prime_q, c^\prime_\ell$}} \\
                & + $q$-loop production &\multirow{3}*{\makecell{}}&\multirow{3}*{\makecell{}} \\
                \bottomrule
    \end{tabular}}
    \caption{eHEFT coefficients for collider searches at NLO order. As laid out in detail in Sec.\,\ref{sec:95GeVres}, additional production processes can contribute, e.g.~VBF, leading to the appearance of further eHEFT coefficients for the left vertex.
    }
    \label{Tab:ListWilsonCoefficients_ColliderSearches}
\end{table}

\section{Additional Effects}
\label{App:additionaleffects}

\subsection{Operators with Chiral Dimension 4}
\label{app:operatorsChDim4}
\begin{figure}
    \centering
    \begin{subfigure}{0.49\textwidth}
    \includegraphics[width=\linewidth]{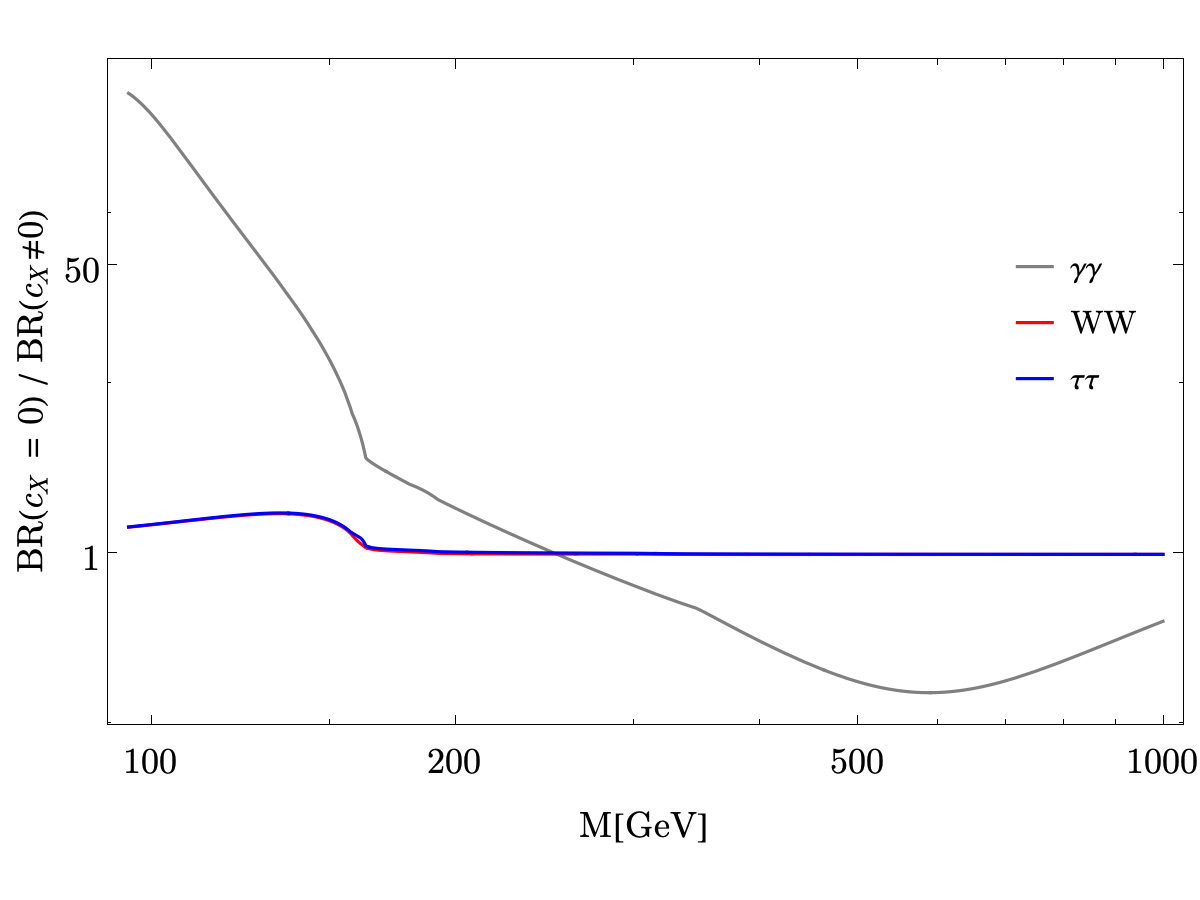}
    \caption{relevant for decays}
    \label{fig:FSTeffectdecay}
    \end{subfigure}
    \hfill
    \begin{subfigure}{0.49\textwidth}
    \includegraphics[width=\linewidth]{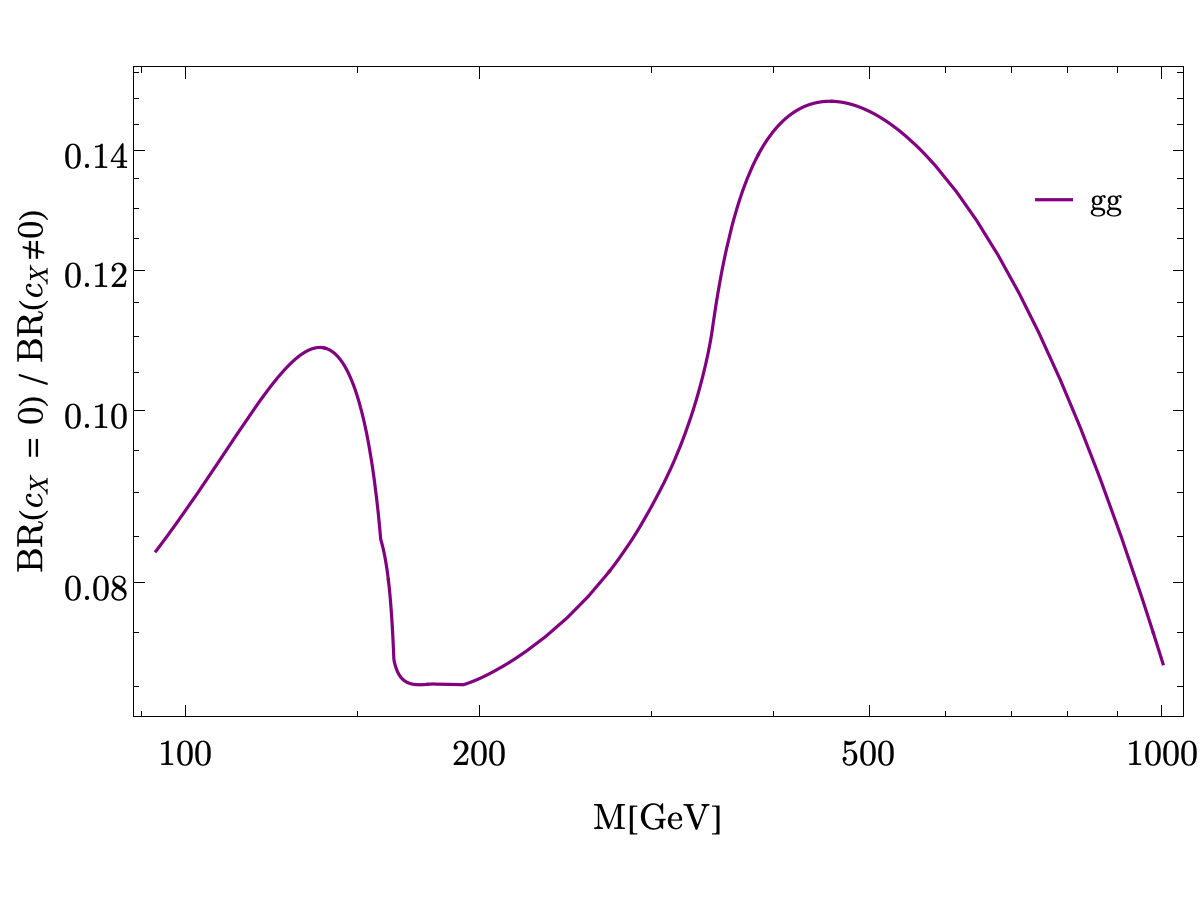}
    \caption{relevant for production}
    \label{fig:FSTeffectprod}
    \end{subfigure}
    \caption{Ratio between expected branching ratio into $\tau \tau, WW, \gamma \gamma, gg$ when setting all $c_X = 0$ or when $c_X \sim 1/\Lambda$. See text for details.}
    \label{fig:FSTeffect}
\end{figure}

Since the operators with chiral dimension four considered in Eq.\,\eqref{Eq:eDMEFT_Lagrangian} are always generated at canonical dimension larger than four, and always carry a loop suppression, we should ask whether their effect might be negligible. Fig.\,\ref{fig:FSTeffect} shows the ratio of branching ratios between $c_X = 0$ and $c_X \sim 1/\Lambda \neq 0$. The branching ratios are calculated with a scalar singlet in mind, i.e. where $c^\prime_f, \kappa^\prime, c_X$ all arise at canonical dimension five. In terms of generating a large effect of $c_X$ this is one of the most favorable models. A higher $SU(2)$ multiplet with vev can have less suppressed $c^\prime_f$ and $\kappa^\prime$ while $c_X$ might have to be generated at higher mass dimension, therefore experiencing an additional suppression. Fig.\,\ref{fig:FSTeffectdecay} shows that the effect the field strength tensor coefficients have on the explored branching ratios is most pronounced for the decay into photons. This makes sense: there is no tree-level contribution to $S \rightarrow \gamma \gamma$ at chiral dimension two. The change in BR ($S \rightarrow WW) $ and BR$( S \rightarrow  \tau \tau$) is simply a consequence of the larger overall decay width due to the boost of $\Gamma(S \rightarrow \gamma \gamma)$. Furthermore, the top and W boson loop contribution to the di-photon signal can partially cancel 
which can be counteracted by the existence of direct couplings of $S$ to the field strength tensors. Fig.\,\ref{fig:FSTeffectprod} shows that, if the coupling to the gluon field strength tensor is generated, it boosts the production of the scalar field. 

\subsection{Decay into Z Bosons}
\label{app:decayZZ}
\begin{figure}
    \centering
    \includegraphics[width=0.5\linewidth]{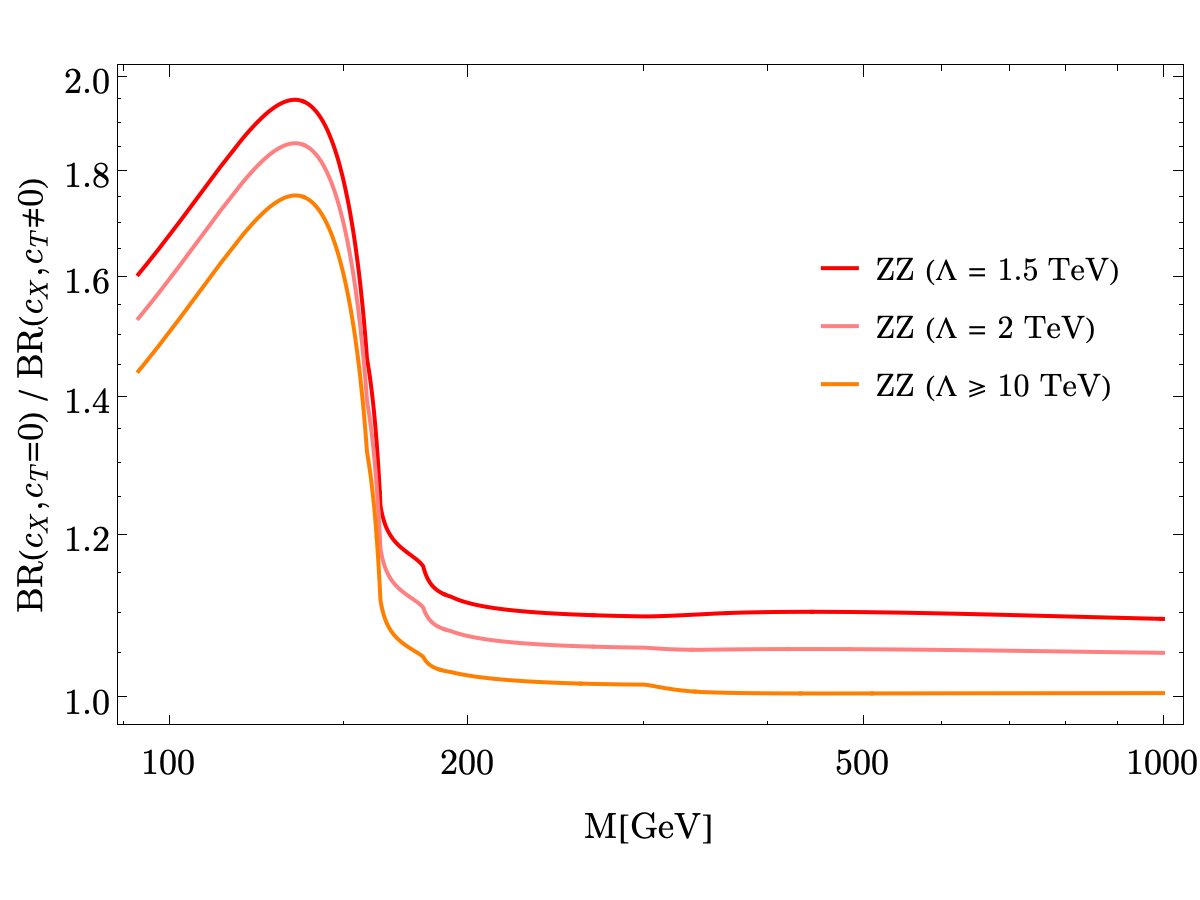}
    \caption{Ratio between expected branching ratios into $ZZ$ for a scalar singlet (model (I)) when setting $\lambda_X = \lambda_{c_T} = 0$ or $\lambda_X \sim \lambda_{c_T} \sim \mathcal{O}(1)$.}
    \label{fig:BrSingletFSTnoT}
\end{figure}

The operator $\mathcal{O}_{c_T}$ modifies the coupling to $Z$ and therefore the decay width for $S \rightarrow ZZ$, see Eq.\,\eqref{eq:decWidthZZ}. The relevant coupling goes as $(\kappa^\prime - 2 c_T^\prime)$ but since $\mathcal{O}_{c_T}$ arises at higher mass dimension than $\mathcal{O}_\kappa$ in most models, there is no strong effect on the branching ratios, as can be seen in Fig.\,\ref{fig:BrSingletFSTnoT}. In general, the larger $\Lambda$, the more negligible $\mathcal{O}_{c_T}$ becomes; for $\Lambda \gtrsim 10$ GeV (orange line), the ratio is not sensitive to whether $\lambda_{c_T} = 0$ or $\neq 0$. A perfect cancellation for $(\kappa^\prime - 2 c_T^\prime)$ were only possible if $\Lambda \lesssim 1$ TeV (using the scaling in Eq.\eqref{eq:scalingSinglet} and assuming $\lambda_i \in [1/3,3]$).

\subsection{Range of Couplings}
\label{app:couplingRange}
\begin{figure}
    \centering
    \begin{subfigure}{0.49\textwidth}
    \includegraphics[width=\linewidth]{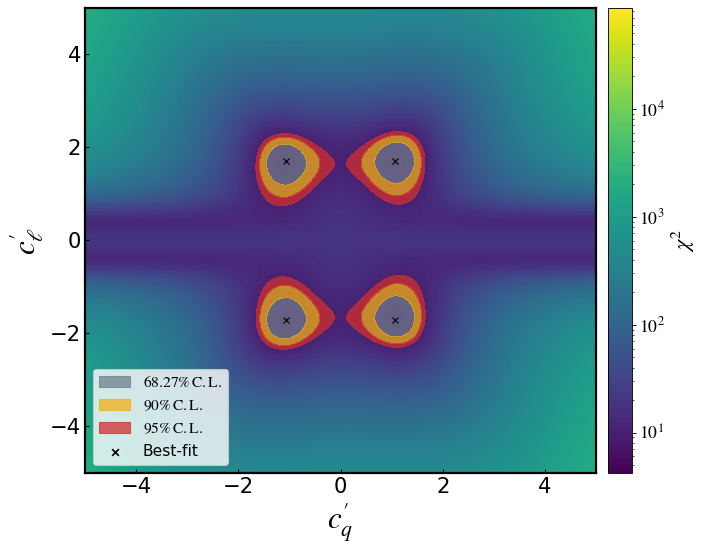}
    \end{subfigure}
    \hfill
    \begin{subfigure}{0.49\textwidth}
    \includegraphics[width=\linewidth]{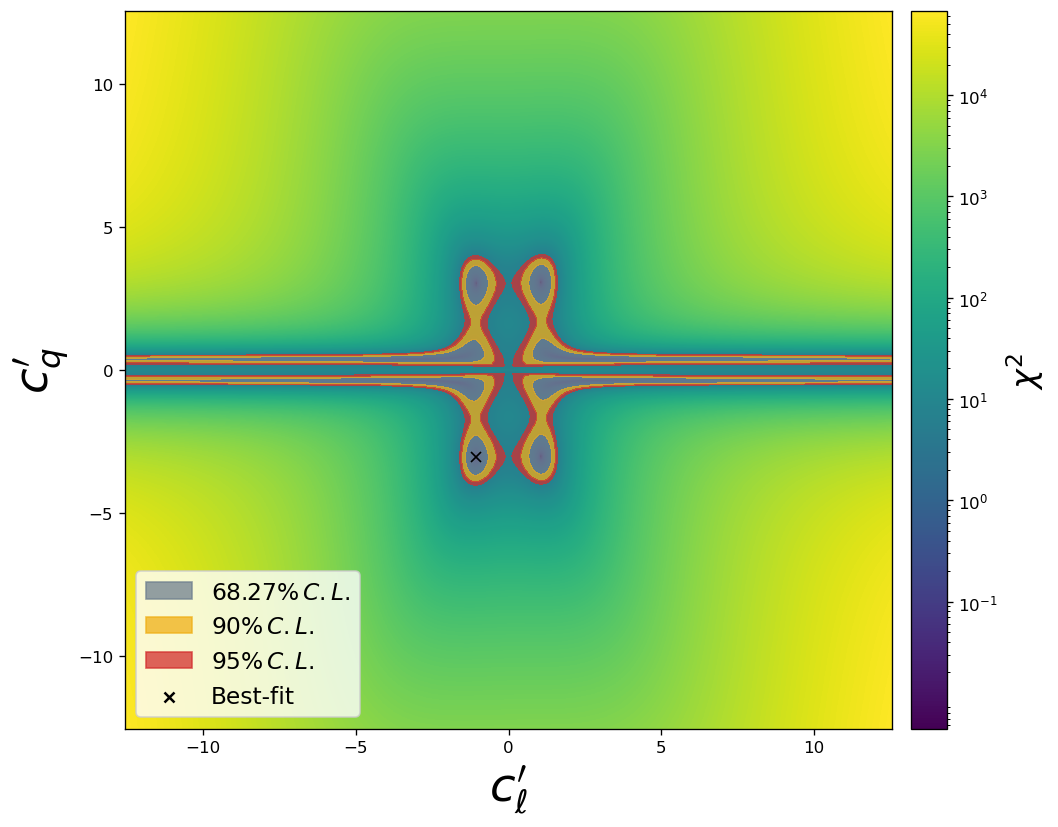}
    \end{subfigure}
    \caption{Exemplary plot showing how the fit is mainly affected by the magnitude of the coefficients and symmetric under sign flips.}
    \label{fig:negCouplings}
\end{figure}

We focus on positive couplings since the predictions for the 95 GeV resonance are mostly sensitive to the absolute value of the coefficients with the exception of the relative sign between $\kappa^\prime$ and $c_f^\prime$. This can be intuitively understood by realizing that most couplings appear squared in the decay widths (Appendix \ref{app:decayWidths}), and we have numerically checked that the mixed terms only modify this behavior in the $\kappa^\prime$–$c_f^\prime$ parameter plane. An example plot is shown in Fig.\,\ref{fig:negCouplings}. For the case of fixed $\kappa^\prime$, the confidence levels are completely symmetric. When marginalizing over the gauge coupling, the minimization of the $\chi^2$ allows for a sign change of $\kappa^\prime$, which restores the symmetry of the plot, as illustrated in the left panel of Fig.\,\ref{fig:negCouplings}. 

\section{Decay Widths}
\label{app:decayWidths}
At NLO order, the relevant decays are described by tree-level contributions from all term in the Lagrangian, as well as the loop-induced decays from the chiral dimension two operators. The two off-shell decays $S \rightarrow WW^\ast$ and $S \rightarrow ZZ^\ast$ are included due to their relevance for the decay of scalars with masses below the $WW/ZZ$ kinematic thresholds. For all decay widths we have crosschecked that the SM results \cite{Djouadi:2005gi} can be recovered by setting $c_W = c_B = c_{WB} =c_G= c^\prime_T = 0$, $\kappa^\prime = 2$, $c_f^\prime = 1$.

\subsection{Tree-level Decays}
By implementing Eq.\,\eqref{Eq:eDMEFT_Lagrangian} in FeynRules \cite{Alloul:2013bka} we determined the tree-level decays relevant for the collider phenomenology. When kinematically allowed, the decay width into two fermions is given by
\begin{equation}
    \Gamma(S\rightarrow f \bar{f} ) = \frac{(y_f c^\prime_f)^2 N_c}{16 \pi} M_S \left(1-4 \frac{m_f^2}{M_S^2}\right)^{3/2} ,
    \label{eq:decWidthfermions}
\end{equation}
where $N_c$ is the color factor, with $N_c^{\text{quark}} = 3$ and $N_c^{\text{lepton}} = 1$. The decay width into gauge bosons is given by
\begin{align}
    \Gamma(S\rightarrow  W W) = &\frac{g^4}{512 \pi} M_S \sqrt{1 - 4 \frac{M_W^2}{M_S^2}}  \nonumber \\
    \times &\left( \frac{(\kappa^\prime)^2}{2} \frac{M_S^2 v^2}{M_W^4} \left( 1 - 4 \frac{M_W^2}{M_S^2} + 12 \frac{M_W^4}{M_S^4}\right)    \right. \nonumber \\
    &+ \frac{6 \kappa^\prime}{\pi^2}  c_W   v \left( 1- 2 \frac{M_W^2}{M_S^2}         \right) \nonumber \\
    &\left. + \frac{(c_W)^2}{\pi^4}\left( 1 - 4 \frac{M_W^2}{M_S^2} + 6 \frac{M_W^4}{M_S^4}\right)  M_S^2 \right) 
    \label{eq:decWidthWW}
\end{align}
and
\begin{align}
    \Gamma(S\rightarrow  Z Z) = &\frac{e^4}{512 \pi^3} M_S \sqrt{1 - 4 \frac{M_Z^2}{M_S^2}}  \nonumber \\
    \times &\left( (\kappa^\prime - 2 c_T^\prime)^2 \frac{\pi^2}{4 \mathfrak{c}_W^4 \mathfrak{s}_W^4} \frac{M_S^2 v^2}{M_Z^4} \left( 1 - 4 \frac{M_Z^2}{M_S^2} + 12 \frac{M_Z^4}{M_S^4}\right) \right. \nonumber \\
    &+ 3 \left(\kappa^\prime - 2 c_T^\prime \right) \left( \frac{c_B}{\mathfrak{c}_W^4} + \frac{c_W}{\mathfrak{s}_W^4} + \frac{c_{WB}^S}{\mathfrak{c}_W^2 \mathfrak{s}_W^2}\right) v \left( 1- 2 \frac{M_Z^2}{M_S^2}  \right)  \nonumber \\
    &+ \left. \frac{1}{2 \pi^2} \left( c_B \mathfrak{s}_W^4 + c_W \mathfrak{c}_W^4 + c_{WB}  \mathfrak{s}_W^2 \mathfrak{c}_W^2 \right)^2  \frac{1}{\mathfrak{c}_W^4 \mathfrak{s}_W^4} \left( 1 - 4 \frac{M_Z^2}{M_S^2} + 6 \frac{M_Z^4}{M_S^4}\right) M_S^2 
    \right) ,
    \label{eq:decWidthZZ}
\end{align}
where $e = \sqrt{4\pi \alpha_{EW}}$, $\mathfrak{c}_W \equiv \cos \theta_W$, $\mathfrak{s}_W \equiv \sin \theta_W$ contain the weak mixing angle $\theta_W$ and the couplings to the field strength tensor terms have mass dimension $[c_W] = [c_B] = [c_{WB}^S]=-1$.  The decay width into photons at tree-level is
\begin{equation}
    \Gamma(S\rightarrow  \gamma\gamma) = \frac{e^4}{1024 \pi^5} M_S^3 \left( c_W + c_B - c_{WB}^S\right)^2,
    \label{eq:decWidthAA}
\end{equation}
and the decay width into $Z\gamma$ is
\begin{equation}
    \Gamma(S\rightarrow  Z \gamma) = \frac{e^4}{512 \pi^5} M_S^3 \left( 1- \frac{M_Z^2}{M_S^2} \right)^3 \frac{1}{\mathfrak{c}_W^2 \mathfrak{s}_W^2} \left(c_W - \left( c_W + c_B\right) \mathfrak{s}_W^2 - \frac{1}{2} c_{WB}^S \left( \mathfrak{c}_W^2 - \mathfrak{s}_W^2 \right)\right)^2.
    \label{eq:decWidthZA}
\end{equation}
Lastly, the decay width into gluons is given by
\begin{equation}
    \Gamma(S\rightarrow  g g) = \frac{g_S^4}{128 \pi^5} M_S^3 \left(c_G\right)^2,
    \label{eq:decWidthgg}
\end{equation}
where $g_S = \sqrt{4 \pi \alpha_S}$ and $[c_G] = -1$, and the decay width into two Higgses is given by
\begin{equation}
    \Gamma(S\rightarrow  h h) = \left( (c_s)_{2,1} \right)^2 \frac{1}{8 \pi M_S} \left( 1 - 4 \frac{M_H^2}{M_S^2}\right)^{1/2},
    \label{eq:decWidthhh}
\end{equation}
where the scalar coupling has mass dimension $[(c_s)_{2,1}] = 1$.

\subsection{Loop-induced Decays from $D_c = 2$ Operators}
The following expressions for the loop-induced decays into $\gamma \gamma, \gamma Z, gg$ are adapted from Sec.\,2.3 in \cite{Djouadi:2005gi}. First, all form factors and auxiliary functions are given by
\begin{equation}
    A_{1/2}(\tau) = 2 [\tau + (\tau -1) f(\tau))]\tau^{-2} ,
\end{equation}
\begin{equation}
    A_1(\tau) = - \left[ 2 \tau^2 + 3 \tau + 3(2 \tau - 1) f(\tau) \right] \tau^{-2},
\end{equation}
\begin{equation}
    A_{1/2}^\prime(\tau^\prime, \lambda) = I_1(\tau^\prime,\lambda) - I_2((\tau^\prime,\lambda)) \,,
\end{equation}
\begin{equation}
    A_1^\prime(\tau^\prime,\lambda) = \mathfrak{c}_W \left( \left( \frac{\mathfrak{s}_W^2}{\mathfrak{c}_W^2} \left( 1 + \frac{2}{\tau^\prime}\right) - \left(5+\frac{2}{\tau^\prime}\right)\right) I_1(\tau^\prime,\lambda) + 4 \left( 3 - \frac{\mathfrak{s}_W^2}{\mathfrak{c}_W^2}\right) I_2(\tau^\prime,\lambda) \right),
\end{equation}
\begin{equation}
    I_1(\tau^\prime,\lambda) = \frac{\tau^\prime \lambda}{2 (\tau^\prime-\lambda)} + \frac{\tau^{\prime2} \lambda^2}{2 (\tau^\prime-\lambda)^2} \left( f(\tau^{\prime-1}) - f(\lambda^{-1})\right) + \frac{\tau^{\prime2} \lambda}{(\tau^\prime-\lambda)^2} \left( g(\tau^{\prime-1}) - g(\lambda^{-1})\right),
\end{equation}
\begin{equation}
    I_2(\tau^\prime,\lambda) = - \frac{\tau^\prime \lambda}{2 (\tau^\prime-\lambda)} \left( f(\tau^{\prime-1}) - f(\lambda^{-1})\right) ,
\end{equation}
\begin{equation}
    f(\tau) = \begin{cases} \arcsin^2 \sqrt{\tau} & \tau \leq 1 \\
    -\frac{1}{4} \left[ \log \frac{1 + \sqrt{1-\tau^{-1}}}{1 - \sqrt{1-\tau^{-1}}} - i \pi\right]^2 & \tau >1
    \end{cases} 
\end{equation}
\begin{equation}
    g(\tau) = \begin{cases} \sqrt{\tau^{-1}-1}\arcsin \sqrt{\tau} & \tau \geq 1 \\
    \frac{\sqrt{1-\tau^{-1}}}{2} \left[ \log \frac{1 + \sqrt{1-\tau^{-1}}}{1 - \sqrt{1-\tau^{-1}}} - i \pi\right] & \tau <1 \end{cases} \,.
\end{equation}
Then, using $G_\mu = 1/(\sqrt{2}v^2)$, $m_f = v y_f/\sqrt{2}$ and replacing the coupling to the Higgs by a coupling to $S$, for the decay width into gluons we obtain
\begin{equation}
    \Gamma_{\text{loop}}(S\rightarrow gg) = \frac{g_s^4}{2304 \pi^5} M_S^3 \left| \sum_q \frac{3}{4} \frac{y_q c_q^\prime}{m_q} A_{1/2}(\tau_q)\right|^2,
\end{equation}
where $\tau_q = M_S^2/4(m^\text{pole}_q)^2$. In a similar fashion, the decay into photons can be derived as
\begin{equation}
    \Gamma_{\text{loop}}(S\rightarrow \gamma \gamma) = \frac{\alpha_{EW}^2}{256 \pi^3} M_S^3 \left| \sum_f \frac{y_f c_f^\prime}{\sqrt{2} m_f} N_c Q_f^2 A_{1/2}(\tau_f) + \frac{\kappa^\prime}{2 v} A_1 (\tau_W) \right|^2,
\end{equation}
where $Q_f$ is the electric charge, and the decay into a photon and a Z as
\begin{align}
    \Gamma_{\text{loop}}(S\rightarrow Z \gamma) = \frac{e^2}{2048 \pi^5} g_W^2 M_S^3 \left( 1-\frac{M_Z^2}{M_S^2}\right)^3 \left| \sum_f \frac{N_f Q_f \hat{v}_f}{\mathfrak{c}_W} \left(\frac{y_f c_f^\prime}{\sqrt{2} m_f}\right) A_{1/2}^\prime (\tau^\prime_f,\lambda_f) + \left(\frac{\kappa^\prime}{2 v} \right) A_1^\prime (\tau^\prime_W,\lambda_W)\right|^2 ,
\end{align}
where $\hat{v}_f = 2 I_f - 4 Q_f \mathfrak{s}_W^2$ with $I_f$ the left-handed weak isospin, $\tau_i^\prime = 4 m_i^2/M_S^2$, $\lambda_i = 4 m_i^2/M_Z^2$. Since some of the considered models have $\kappa^\prime=0$ and therefore no direct coupling to W or Z gauge bosons, the loop-induced decays into the EW gauge bosons can become important. They become
\begin{align}
    \Gamma_{\text{loop}}(S\rightarrow Z Z) = \frac{e^4}{8192 \pi^5} \frac{1}{\mathfrak{c}_W^4 \mathfrak{s}_W^4} M_S^3 \sqrt{1-4\frac{M_Z^2}{M_S^2}} \left(1 - 4 \frac{M_Z^2}{M_S^2} +6 \frac{M_Z^4}{M_S^4} \right) \left| \sum_f \frac{4}{3} \frac{y_f c^\prime_f}{m_f}A_{1/2} (\tau_f) \left( \mathfrak{c}_W^4 + \mathfrak{s}_W^4\right)\right|^2 
\end{align}
and
\begin{align}
    \Gamma_{\text{loop}}(S\rightarrow WW) = \frac{e^4}{4096 \pi^5} \frac{1}{\mathfrak{s}_W^4} M_S^3 \sqrt{1-4\frac{M_W^2}{M_S^2}} \left(1 - 4 \frac{M_W^2}{M_S^2} +6 \frac{M_W^4}{M_S^4} \right) \left| \sum_f \frac{4}{3} \frac{y_f c^\prime_f}{m_f}A_{1/2} (\tau_f)\right|^2 \, ,
\end{align}
and we neglect the Higgs-induced loop contribution. When the loop- and tree-level decays are comparable in size, as is the case for $S \rightarrow gg, \gamma \gamma, Z \gamma$, we furthermore consider interference effects arising from $\Gamma \propto|\mathcal{M}_{\text{loop}} + \mathcal{M}_{\text{tree}}|^2 \supset 2 \mathcal{M}_{\text{loop}} \mathcal{M}_{\text{tree}} $. For $S \rightarrow WW, ZZ$ the tree-level decay dominates if present, thus allowing to safely neglect the (then small) loop corrections to the decay widths.

\subsection{Off-shell Decays into W and Z Bosons}
For scalar masses below $2 m_W$ / $2 m_Z$, where $S \rightarrow V V$ is kinematically forbidden, the off-shell decay $S \rightarrow V V^\ast \rightarrow V f \bar{f}$ becomes relevant, leading to a three-body final state. As above, we first list auxiliary functions
\begin{align}
    R_T(x) = & \frac{3 \left(1-8x + 20 x^2 \right)}{\left( 4x-1\right)^{1/2}} \arccos \frac{3x-1}{2x^{3/2}} \nonumber \\
    &- \frac{1-x}{2x} \left( 2 - 13 x + 47 x^2\right) \nonumber \\
    &- \frac{3}{2} \left( 1- 6 x + 4 x^2\right) \log x,
\end{align}
\begin{align}
    R_T^\prime(x) = & - \frac{  28x^{5/2}-16x^{3/2} + 2x^{1/2}}{\left( 4x-1\right)^{1/2}} \arccos \frac{3x-1}{2x^{3/2}}  \nonumber \\
    &- \left( 9x^{5/2} -14x^{3/2}+5x^{1/2} \right) \nonumber \\
    & +  \left(2x^{5/2} - 6 x^{3/2} + x^{1/2} \right) \log x ,
\end{align}
\begin{align}
    R_T^{\prime\prime} (x)= & \frac{6 \left(54x^3 - 40x^2 + 11 x -1 \right)}{\left( 4x-1\right)^{1/2}} \arccos \frac{3x-1}{2x^{3/2}} \nonumber \\
    &+ \left( 89 x^3 - 171x^2 + 99x - 17 \right) \nonumber \\
    &- 3 \left(6x^3 -30x^2 + 9x - 1 \right) \log x,
\end{align}
where $x = M_V^2/M_S^2$, to give the decay widths in a compact form as
\begin{align}
    \Gamma (S \rightarrow W W^\ast) = \frac{M_S}{2048 \pi^3} g_W^4 &\left(3 (\kappa^{\prime})^2  
    R_T(x) \right. 
    + \frac{9 M_S}{\pi^2} g_W R_T^\prime(x) \kappa^\prime  c_W \nonumber \\
    &+ \left. \frac{M_S^2}{4 \pi^4} g_W^2 R_T^{\prime \prime}(x)  c_W^2 \right) ,
\end{align}
\begin{align}
    \Gamma (S \rightarrow Z Z^\ast) = \frac{M_S}{2048 \pi^3 } \frac{g_W^4}{\mathfrak{c}_W^4 } \delta_Z &\Biggl\{ 3 (\kappa^{\prime} - 2 c_T^\prime)^2 R_T(x) + \frac{9 M_S}{\pi^2 } \frac{g_W}{\mathfrak{c}_W } R_T^\prime (x) (\kappa^{\prime} - 2 c_T^\prime) \left[ c_W \mathfrak{s}_W^4 + c_B \mathfrak{c}_W^4 + c_{WB} \mathfrak{s}_W^2 \mathfrak{c}_W^2\right] \nonumber \\
    &+ \frac{M_S^2}{4 \pi^4} \frac{g_W^2}{\mathfrak{c}_W^2} R_T^{\prime \prime}(x) \left[ c_W \mathfrak{s}_W^4 + c_B \mathfrak{c}_W^4 + c_{WB} \mathfrak{s}_W^2 \mathfrak{c}_W^2\right]^2 \Biggl\}, \nonumber \\
\end{align}
where
\begin{equation}
     \delta_Z = \frac{40}{27} \mathfrak{s}_W^4 - \frac{10}{9} \mathfrak{s}_W^2 + \frac{7}{12}.
\end{equation}

\end{appendix}
\bibliographystyle{JHEP} 
\apptocmd{\thebibliography}{\justifying}{}{} 
\bibliography{eDMEFT2}
\end{document}